\documentclass[preprintnumbers,article,amsmath,amssymb,floatfix,10pt,prd,twocolumn,
superscriptaddress,nofootinbib]{revtex4-2}
\usepackage{bm}
\usepackage{amsfonts}
\usepackage{latexsym}
\usepackage[latin1]{inputenc}
\usepackage{graphicx}
\usepackage{amsmath}
\usepackage{palatino}
\usepackage{mathpazo}
\usepackage{textcomp}
\usepackage{float}
\usepackage{booktabs}
\usepackage{dcolumn}
\usepackage{ragged2e}
\usepackage{hyperref}
\hypersetup{colorlinks,citecolor=blue}
\hypersetup{colorlinks=true,linkcolor=red,filecolor=magenta,    urlcolor=blue}
\usepackage{amsmath}
\usepackage{xcolor}
\usepackage{orcidlink}
\usepackage{epsfig}
\usepackage{subfigure}
\usepackage{commath}

\def\jnl@style{\it}
\def\aaref@jnl#1{{\jnl@style#1}}

\def\aaref@jnl#1{{\jnl@style#1}}

\def\aj{\aaref@jnl{AJ}}                   
\def\apj{\aaref@jnl{ApJ}}                 
\def\apjl{\aaref@jnl{ApJ}}                
\def\apjs{\aaref@jnl{ApJS}}               
\def\apss{\aaref@jnl{Ap\&SS}}             
\def\aap{\aaref@jnl{A\&A}}                
\def\aapr{\aaref@jnl{A\&A~Rev.}}          
\def\aaps{\aaref@jnl{A\&AS}}              
\def\mnras{\aaref@jnl{Mon.~Not.~Roy.~Astron.~Soc.}}             
\def\prd{\aaref@jnl{Phys.~Rev.~D}}        
\def\prc{\aaref@jnl{Phys.~Rev.~C}}  
\def\prl{\aaref@jnl{Phys.~Rev.~Lett.}}    
\def\qjras{\aaref@jnl{QJRAS}}             
\def\skytel{\aaref@jnl{S\&T}}             
\def\ssr{\aaref@jnl{Space~Sci.~Rev.}}     
\def\zap{\aaref@jnl{ZAp}}                 
\def\nat{\aaref@jnl{Nature}}              
\def\aplett{\aaref@jnl{Astrophys.~Lett.}} 
\def\apspr{\aaref@jnl{Astrophys.~Space~Phys.~Res.}} 
\def\physrep{\aaref@jnl{Phys.~Rep.}}      
\def\physscr{\aaref@jnl{Phys.~Scr}}       
\def\commat{\aaref@jnl{Comm.~Math.~Phys.}}              
\def\science{\aaref@jnl{Science}}               
\def\cqg{\aaref@jnl{Classical Quant.~Grav.}}            
\def\jpcs{\aaref@jnl{JPCS}}                                     
\def\ijmpd{\aaref@jnl{Int.~J.~Mod.~Phys.~D}}                    
\def\grg{\aaref@jnl{Gen.~Relat.~Gravit.}}               
\def\rpp{\aaref@jnl{Rep.~Prog.~Phys.}}          
\def\npa{\aaref@jnl{Nucl.~Phys.~A}}        
\def\lrr{\aaref@jnl{Living Rev.~Rel.}}                   
\def\jcap{\aaref@jnl{J.~Cosmology Astropart.~Phys.}}    
\def\rmp{\aaref@jnl{Rev.~Mod.~Phys.}}   
\def\epjc{\aaref@jnl{Eur.~Phys.~J.~C}} 
\def\plb{\aaref@jnl{~Phy.~Lett.~B}} 
\def\mpla{\aaref@jnl{Mod.~Phy.~Lett.~A}} 
\def\arxiv{\aaref@jnl{arxiv.org}}

\allowdisplaybreaks[1]

\begin{document}
\color{black}       
\title{Possibility of the Traversable Wormholes in the Galactic Halos within $4D$ Einstein-Gauss-Bonnet Gravity}

\author{Zinnat Hassan\orcidlink{0000-0002-6608-2075}}
\email{zinnathassan980@gmail.com}
\affiliation{Department of Mathematics, Birla Institute of Technology and
Science-Pilani,\\ Hyderabad Campus, Hyderabad-500078, India.}

\author{P.K. Sahoo\orcidlink{0000-0003-2130-8832}}
\email{pksahoo@hyderabad.bits-pilani.ac.in}
\affiliation{Department of Mathematics, Birla Institute of Technology and
Science-Pilani,\\ Hyderabad Campus, Hyderabad-500078, India.}

%
\date{\today}
\begin{abstract}
Recently, there has been significant interest regarding the regularization of a $D\rightarrow 4$ limit of Einstein-Gauss-Bonnet (EGB) gravity. This regularization involves re-scaling the Gauss-Bonnet (GB) coupling constant as $\alpha/(D-4)$, which bypasses Lovelock's theorem and avoids Ostrogradsky instability. A noteworthy observation is that the maximally or spherically symmetric solutions for all the regularized gravities coincide in the $4D$ scenario. Considering this, we investigate the wormhole solutions in the galactic halos based on three different choices of dark matter (DM) profiles, such as Universal Rotation Curve, Navarro-Frenk-White, and Scalar Field Dark Matter with the framework of $4D$ EGB gravity. Also, the Karmarkar condition was used to find the exact solutions for the shape functions under different non-constant redshift functions. We discussed the energy conditions for each DM profile and noticed the influence of GB coefficient $\alpha$ in violating energy conditions, especially null energy conditions. Further, some physical features of wormholes, viz. complexity factor, active gravitational mass, total gravitational energy, and embedding diagrams, have been explored.
\end{abstract}

\maketitle


\textbf{Keywords:} Galactic halos, dark matter, wormhole solutions, $4D$ EGB gravity, energy conditions.
\section{Introduction}\label{sec1}
\indent In 1915, Albert Einstein published his seminal work, which revolutionized our understanding of the fundamental nature of gravity by linking it to the geometry of space-time. Over the subsequent years, extensive experimental and observational validation \cite{Leonard} has solidified the significance of this theory, making it an indispensable tool for researchers worldwide in their pursuit of a deeper comprehension of the cosmos. This groundbreaking theory has given rise to many extraordinary phenomena, including gravitational lensing, the formation of black holes, the detection of gravitational waves, and the concept of wormholes. One of the particular interests among these phenomena is the theoretical existence of wormholes, an alternative solution within Einstein's Field Equations (EFE) that conceivably bridges the gap between distinct regions of space-time.\\
\indent Interest in wormhole space-times traces its origins back to 1916 when Flamm \cite{Flamm} initially postulated the existence of a \textit{tunnel structure} within the Schwarzschild solution, suggesting it to be a potential wormhole. Later, Einstein and Rosen \cite{Rosen} introduced the concept of a ``bridge structure" that links two exterior regions of a Schwarzschild black hole space-time, thereby giving rise to an inter-universe connection. 
Wheeler \cite{Wheeler1,Wheeler2} coined the term ``wormhole" to describe microscopic charge-carrying wormholes. They demonstrated that if a wormhole somehow attempted to open, it would close so rapidly that even a single photon might travel through it, thereby sustaining Einsteinian causality. A detailed discussion of the possibility of a wormhole was presented in \cite{Riaz0}. The potential existence of wormholes in the galactic halo region was discussed in \cite{Ovgun0}, and gravitational lensing by wormholes was explored in \cite{Riaz1,Riaz2,Riaz3,Riaz4,Riaz5,Riaz6}. For further interesting works on wormhole geometry, one may check some Refs. \cite{Maldacena1,Maldacena2,Maldacena3,Maldacena4,Maldacena5}.\\
\indent In General Relativity (GR), a key element in understanding wormholes involves violating energy conditions \cite{Thorne/1988}, and the matter associated with violating energy conditions is termed as exotic matter. Since minimizing the use of exotic matter poses a primary challenge in wormhole theory, various approaches have been explored in the literature. These include the cut-and-paste method \cite{Dadhich1,Dadhich4} and the investigation of wormholes with variable equations of state (EoS) \cite{Garattini1,Garattini2}. 
Simultaneously, exploring modified theories of gravity has provided a fresh perspective on wormhole studies. Numerous modified gravity theories have been employed to delve into wormhole theory, where the additional terms in the modified field equations facilitate solutions aligning with the satisfaction of energy conditions by ordinary matter. 
Readers can check some works on wormhole geometry in different modified theories include $f(R)$ gravity \cite{Karakasis,Golchin,Eid,Goswami}, $f(R,T)$ gravity \cite{Bhatti,Chanda,Rosa}, braneworld scenarios \cite{Wong1,Wong3,Wong4}, Rastall theory \cite{Ovgun1,Ovgun2,Ovgun3}, $f(Q)$ gravity \cite{Mustafa,Hassan1,Banerjee1,Hassan2}, and $f(Q,T)$ gravity \cite{Tayde1,Tayde2,Tayde3}.\\
\indent In string theories, particularly in their low-energy limits, effective field theories of gravity emerge. In these theories, the Lagrangian incorporates terms of quadratic and higher orders in the curvature alongside the conventional scalar curvature term \cite{Zwiebach1,Zwiebach3,Zwiebach4}. Notably, the gravitational action can undergo modification to encompass quadratic curvature correction terms while maintaining the equations of motion in the second order. This modification is permissible as long as the quadratic terms manifest in specific combinations corresponding to the GB invariants denoted by \cite{Lanczos}
\begin{equation}\label{a1}
\mathcal{G}=R^2-4R_{\alpha\nu}\,R^{\alpha\nu}+R_{\alpha\nu\delta\eta}\,R^{\alpha\nu\delta\eta}.
\end{equation}
Such a gravitational theory is designated as $D\geq 5$ dimensional EGB gravity theory, with $(D-4)$ extra dimensions playing a pivotal role. 
The EGB gravity constitutes a specific instance within the broader framework of the Lovelock theory of gravitation \cite{Lovelock1}. Significantly, the equations of motion associated with EGB gravity exhibit no more than two metric derivatives, ensuring the absence of ghost-related issues \cite{Lovelock2}. 
However, it is crucial to note that the GB term \eqref{a1} behaves as a topological invariant in four dimensions, contributing proportionally to $(D-4)$ in all components of Einstein's equations. Consequently, this term does not contribute to the equations of motion, necessitating $ D\geq 5$ to manifest non-trivial gravitational dynamics.\\
\indent Nevertheless, Glavan and Lin \cite{Lin1} demonstrated a significant impact on gravitational dynamics in $D = 4$ could be achieved by re-scaling the GB coupling constant in the EGB gravity action, expressed as 
\begin{equation}\label{a2}
\alpha \rightarrow \frac{\alpha}{D-4}.
\end{equation}
A consistent theory of $D\rightarrow 4$ EGB gravity, maintaining invariance under spatial diffeomorphism while breaking the time diffeomorphism, has been discussed in \cite{Aoki1,Aoki2}. The 4D EGB gravity has garnered substantial attention and is undergoing extensive examination \cite{Lin2,Lin4,Lin5,Lin6}. 
A cloud of strings in 4D EGB was also explored in \cite{Lin8}. The exploration of non-static Vaidya-like spherical radiating black hole solutions in 4D EGB gravity has been reported \cite{Lin9}, and further black hole solutions can be found in \cite{Doneva1,Doneva2,Doneva3,Doneva4,Doneva5}. Furthermore, in Ref. \cite{Guo}, the authors discussed the photon geodesics and the influence of the GB coupling parameter on the shadow of the 4D EGB black hole. A detailed review of this topic has been discussed in \cite{Mulryne}. 
Wormhole solutions were also investigated using the same formalism in $4D$ EGB gravity. Jusufi et al. \cite{Jusufi1} discussed wormhole solutions in $4D$ EGB gravity for isotropic and anisotropic matter sources. Moreover, the stability of thin-shell wormholes \cite{Jusufi2} and Yukawa-Casimir wormholes \cite{Jusufi3} have been investigated. However, wormholes in the region of galactic dark matter halos have yet to be explored, which motivates us to study this particular gap.\\
\indent Recent revelations about the composition of the Universe have brought to light the fact that a mere 5\% of the total matter and energy consists of the visible planets, stars, and ordinary matter, with the remainder being dominated by dark matter and dark energy of which 27\% is attributed to dark matter. Unlike regular baryonic matter, this dark matter does not interact with photons but exerts its influence solely through gravitational forces. In 1933, astronomer Zwicky was the first to ascertain the presence of dark matter within a distant galaxy cluster \cite{Zwicky1,Zwicky2}. Oort \cite{Oort1,Oort2} initially proposed the existence of dark matter in the Milky Way galaxy and subsequently confirmed through observational evidence by Diemand and Springel \cite{Diemand,Springel}. 
Dark matter plays a crucial role in the hypothetical construction of wormholes, with the density profile of dark matter proposed by researchers playing a pivotal role in the theoretical development of a fully stable and traversable wormhole. In Ref. \cite{Rahaman6}, the authors have provided evidence indicating the role of DM in sustaining the wormhole geometry within the outer reaches of the galactic halo. Sarkar et al. \cite{Sarkar2} have demonstrated the presence of wormholes within the isothermal galactic halo, supported by the presence of DM. Additionally, Kuhfitting \cite{Rahaman8} has examined the gravitational lensing phenomena associated with wormholes within the galactic halo region.\\
\indent In recent years, there has been a remarkable surge in the number of exact solutions describing compact objects within classical GR and modified gravity theories. This surge has pushed the search for solutions to the field equations into the forefront of mainstream astrophysics, particularly evidenced in the expansion of realistic stellar models over the last decade. 
Utilization of the Karmarkar condition has played a pivotal role in generating compact objects, enabling the differentiation between the radial and transverse stresses at various points within the stellar fluid \cite{Ratanpal1,Ratanpal2}. In the context of classical GR, the Karmarkar condition is readily integrated to establish a relationship between the two metric potentials, effectively simplifying the task of finding exact solutions to the field equations into a single generating function \cite{Bhar1,Bhar2}. The authors of Ref. \cite{Hansraj} presented a convincing discussion regarding the nonexistence of conformally flat-charged isotropic fluid spheres falling under the category of embedding Class One. Also, the Karmarkar condition has been extended to encompass time-dependent systems, enabling its utilization in scenarios involving the modeling of shear-free, dissipative collapse \cite{Naidu1,Naidu2,Naidu3}. Moreover, in \cite{Maurya4}, anisotropic stars have been studied via Karmarkar condition in Brans-Dicke gravity and confirm that their solutions describe compact objects such as PSR J1903+327; Cen X-3; EXO 1785-248 and LMC X-4 to an excellent approximation. Furthermore, the Karmarkar condition has been established to be a valuable tool in exploring other compact objects, including wormholes.\\
\indent Motivated by the above discussions, we have investigated wormhole solutions under different DM halo models by employing the Karmarkar condition in the context of $4D$ Gauss-Bonet gravity. The structure of this paper is outlined as follows: In section \ref{sec2}, we have provided the basic formalism of $4D$ EGB gravity and corresponding wormhole field equations under this gravity. The Karmarkar condition and the DM halo profiles have been extensively discussed in section \ref{sec3} and \ref{sec4}, respectively. Moreover, section \ref{sec5} discusses wormhole solutions for different redshift functions under each DM profile. Further, some physical features of wormholes have been studied in section \ref{sec6}. Finally, we conclude our finding in section \ref{sec7}.

\section{Basic formalisms of $4D$ EGB gravity and wormhole field equations}\label{sec2}
\subsection{Basic formalisms of $4D$ EGB gravity}
In the $D$ dimensions, the EGB gravity action incorporating a rescaled GB coupling $\alpha/(D-4)$ to restore dimensional regularization results in the expression \cite{Lin1}
\begin{equation}\label{2a1}
\mathcal{I}=\frac{1}{16 \pi}\int d^D\,x\sqrt{-g}\left(R+\frac{\alpha}{D-4} \mathcal{G}\right)+\mathcal{S}_{matter},
\end{equation}
where $R$ represents the Ricci scalar,  and $\alpha$ is the GB coefficient with dimension $[\textit{Length}]^2$. The GB invariant is described by $\mathcal{G}$, and its expression is given in Eq. \eqref{a1}. The $4D$ EGB theory is defined by considering the limit $D\rightarrow 4$ at the level of equations of motion rather than in the action. Consequently, in this framework, the GB term plays a non-trivial role in gravitational dynamics \cite{Lin1}.\\
Now, by varying the action \eqref{2a1} w.r.t the metric tensor, one can find the field equations
\begin{equation}\label{2a3}
G_{\mu\nu}+\frac{\alpha}{D -4}\mathcal{H}_{\mu\nu}=8\pi\mathcal{T}_{\mu\nu},
\end{equation}
where $G_{\alpha\nu}$, $\mathcal{T}_{\alpha\nu}$ and $\mathcal{H}_{\alpha\nu}$ are the Einstein tensor, energy-momentum tensor and  the GB tensor, respectively which are defined by
\begin{equation}
G_{\mu\nu}=R_{\mu\nu}-\frac{1}{2} R g_{\mu\nu},
\end{equation}
\begin{equation}
\mathcal{T}_{\mu\nu}=-\frac{2}{\sqrt{-g}}\frac{\delta\left(\sqrt{-g}\,\mathcal{S}_m\right)}{\delta g^{\mu\nu}},
\end{equation}
and
\begin{multline}
\mathcal{H}_{\mu\nu}=-\frac{1}{2}\mathcal{G}g_{\mu\nu}+2\left[R\,R_{\mu\nu}-2R_{\mu\sigma}R^{\sigma}_{\,\,\nu}-2R_{\mu\rho\nu\sigma}\,R^{\rho\sigma}\right.\\\left.
-2R_{\mu\sigma b c}\,R^{\sigma b c}_{\,\,\,\,\,\,\,\,\nu}\right],
\end{multline}
where $R_{\mu\nu}$ represents the Ricci tensor and $R_{\mu\rho\nu\sigma}$ is the Riemann tensor. The GB terms are generally characterized as total derivatives in $4D$ space-time, thereby not contributing to the field equations. Nevertheless, when the coupling constant $\alpha/(D-4)$ undergoes re-scaling and maximally symmetric space-times with a curvature scale $\mathcal{K}$, we obtain
\begin{equation}\label{a3}
\frac{g_{\mu \sigma}}{\sqrt{-g}}\frac{\partial \mathcal{G}}{\partial g_{\nu \sigma}}=\frac{\alpha (D-2) (D-3)}{2 (D-1)} \mathcal{K}^2 \delta^{\nu}_{\mu}.
\end{equation}
Evidently, the GB action variation does not vanish in $D = 4$ due to the influence of the re-scaled coupling constant \cite{Lin1}.\\
To generate wormhole solutions within the framework of $4D$ EGB, we employ the regularization process outlined in \cite{Lin1}. Notably, the $4D$ spherical solutions derived in \cite{Lin1} align with those obtained in various other regularized theories \cite{Hennigar1,Hennigar2,Hennigar3}.
\subsection{Wormhole field equations}
In this analysis, we focus on a $D$-dimensional spacetime where the two-sphere \cite{Thorne/1988} is substituted with an $(D-2)$ sphere, as described by the following line element
\begin{equation}\label{2c1}
ds^2=-e^{2\phi(r)}dt^2+\left(1-\frac{b(r)}{r}\right)^{-1}dr^2+r^2d\Omega^2_{D-2},
\end{equation}
where $$d\Omega^2_{D-2}=d\theta_1^2+\sum\limits_{i=2}^{D-2}\prod\limits_{j=1}^{i-2} \sin^2\theta_j d\theta_i^2.$$
$\phi(r)$ represents the redshift function, and to bypass the event horizon issue, the redshift function must be finite everywhere. $b(r)$ is the shape function; it determines the shape of the wormholes. It is crucial to note that $b(r)$ must adhere to the boundary condition $b(r = r_0) = r_0$ at the throat, where $r_0$ is such that $r_0 \leq r \leq \infty$. To assure the traversability of a wormhole, b(r) should meet the flaring-out condition, which is deduced from the embedding calculation and expressed as $(b-b'r)/b^2>0$ \cite{Thorne/1988}. Alternatively, this condition can be compactly stated as $b(r_0) < 1$ at the throat, where $r = r_0$. Besides, the condition $1-\frac{b(r)}{r}>0$ is enforced.\\
Further, we consider the diagonal energy-momentum tensor for an anisotropic fluid, which can be read as
\begin{equation}\label{11}
\mathcal{T}_{\mu}^{\nu}=\text{diag}[-\rho(r),p_r(r),p_t(r),p_t(r)],
\end{equation}
where $\rho(r)$ is the energy density. $p_r(r)$ and $p_t(r)$ represent the radial pressure and tangential pressure, respectively.\\
Thus, the gravitational field equation \eqref{2a3} for the metric \eqref{2c1} in the limit $D\rightarrow 4$ provides the following relations \cite{Jusufi1}
\begin{equation}\label{2c3}
8\pi\rho(r)=\frac{\alpha b(r)}{r^6}\left(2rb'(r)-3b(r)\right)+\frac{b'(r)}{r^2},
\end{equation}
\begin{multline}\label{2c4}
8\pi p_r(r)=\frac{\alpha b(r)}{r^6}\left[4 r\phi'(r)\left(r-b(r)\right)+b(r)\right]\\
+\frac{2\phi'(r)\left(r-b(r)\right)}{r^2}-\frac{b(r)}{r^3},
\end{multline}
\begin{multline}\label{3c5}
8\pi p_t(r)=\left(1-\frac{b(r)}{r}\right)\left[\left(\phi''(r)+(\phi'(r))^2\right)\right.\\\left.
\times \left(1+\frac{4\alpha b(r)}{r^3}\right)
+\frac{1}{r}\left(\phi'(r)-\frac{r b'(r)-b(r)}{2 r (r-b(r)}\right)\right.\\\left.
\times \left(1-\frac{2\alpha b(r)}{r^3}\right)
-\left(\frac{\left(r b'(r)-b(r)\right)\phi'(r)}{2 r (r-b(r)}\right) \right.\\\left.
\times \left(1-\frac{8\alpha}{r^2}+\frac{12 \alpha b(r)}{r^3}\right)\right]-\frac{2\alpha b^2(r)}{r^6},
\end{multline}
where the prime signifies a derivative w.r.t. radial coordinate $r$.\\
Let me briefly elaborate on classical energy conditions derived from the Raychaudhuri equations. Given our focus on anisotropic fluid matter distribution, the energy conditions derived from standard General Relativity take the form\\
$\bullet$ Null energy condition (NEC) if $\rho+P_r\geq0$, \quad  $\rho+P_t\geq0$.\\
$\bullet$ Weak energy conditions (WEC) if $\rho\geq0$, \quad $\rho+P_r\geq0$, \quad  $\rho+P_t\geq0$.\\
$\bullet$ Dominant energy conditions (DEC) if $\rho-\abs{p_r}\geq0$, \quad  $\rho-\abs{p_t} \geq0$.\\
$\bullet$ Strong energy conditions (SEC) if $\rho+P_r+2P_t\geq0$.\\
Considering these, we will examine these energy conditions by defining\\
$NEC1=\rho+P_r$, \quad $NEC2=\rho+P_t$, \quad $DEC1=\rho-\mid P_r \mid$, \quad $DEC2=\rho-\mid P_t \mid$, \quad $SEC=\rho+P_r+2P_t$\\
for this investigation.
\section{Embedding class-1 space-time}\label{sec3}
In the present work, our main focus is to develop a wormhole shape function using the Karmarkar conditions that describe wormhole geometry. For this purpose, we assume static spherically symmetric space-time defined as:
\begin{equation}\label{2b1}
ds^2=-e^{\xi(r)}dt^2+e^{\lambda(r)}dr^2+r^2(d\theta^2+\sin^2\theta d\phi^2),
\end{equation}
where $\xi$ and $\lambda$ are function of radial coordinate $r$ only. According to the above line element, the non-vanishing
covariant Riemannian curvature components are
\begin{multline}\label{2b2}
\,\,\,\,\,\,\,\,R_{1212}=R_{2121},\,\,\,R_{1221}=R_{2112},\,\,\,R_{1313}=R_{3131},\\
R_{1331}=R_{3113},\,\,\,R_{2323}=R_{3232},\,\,\,R_{2332}=R_{3223},\\
R_{1414}=R_{4141},\,\,\,R_{4114}=R_{1441},\,\,\,R_{4242}=R_{2424},\\
R_{4224}=R_{2442},\,\,\,R_{4343}=R_{3434},\,\,\,R_{4334}=R_{3443}.
\end{multline}
The concept of embedding $n$-dimensional space $V_n$ within a pseudo-Euclidean space $E_n$ has been a subject of significant interest, as highlighted in the works of Eisland \cite{Eiesland} and Eisenhart \cite{Eisenhart1}. When an $n$-dimensional space $V_n$ can be isometrically immersed in an $(n + m)$-dimensional space, where $m$ represents the minimum number of additional dimensions, $V_n$ is considered to possess $m$-Class embedding, in that context, metric described in \eqref{2b1} typically yields a four-dimensional spherically symmetric space-time, placing it within Class two, denoting $m = 2$ and its embedding in a six-dimensional pseudo-Euclidean space. On a different note, it's worth mentioning that it is possible to devise a parametrization that allows the space-time outlined in equation \eqref{2b1} to be incorporated into a five-dimensional pseudo-Euclidean space, corresponding to Class $m = 1$, commonly referred to as embedding Class one \cite{Eiesland,Eisenhart1,Karmarkar1}. In order for a spherically symmetric space-time, whether static or non-static, to be classified as Class-one, the system must adhere to the following requisite and appropriate conditions:
\begin{itemize}
\item the Gauss equation:
\begin{eqnarray}\label{eqcls1.1}
\mathcal{R}_{mnpq}=2\,\epsilon\,{b_{m\,[p}}{b_{q]n}}\,,
\end{eqnarray}
\item the Codazzi equation:
\begin{eqnarray}\label{eqcls1.2}
b_{m\left[n;p\right]}-{\Gamma}^q_{\left[n\,p\right]}\,b_{mq}+{{\Gamma}^q_{m}}\,{}_{[n}\,b_{p]q}=0.
\end{eqnarray}
\end{itemize}
Here, we consider the case where $\epsilon=\pm1$ and square brackets denote antisymmetrization. The coefficients of the second differential form are represented by $b_{mn}$. By utilizing Eqs.~(\ref{eqcls1.1}) and ~(\ref{eqcls1.2}) and applying the prescribed mathematical procedure, we can calculate the Karmarkar condition as follows:
\begin{equation}\label{2b5}
R_{1414}=\frac{R_{1224}R_{1334}+R_{1212}R_{3434}}{R_{2323}},
\end{equation}
with $R_{2323}\neq0$ \cite{Pandey1}. The form of space-time satisfying the Karmarkar condition is known as embedding class-I. By substituting the non-zero
components of Riemann curvature \eqref{2b2} in the Karmarkar relation \eqref{2b5}, we get the following differential equation:
\begin{equation}\label{2b6}
\frac{\xi'(r) \lambda'(r)}{1-e^{\lambda(r)}}=\xi'(r)\lambda'(r)-\xi'^2(r)-2\xi^{''},
\end{equation}
where $e^{\lambda(r)}\neq1$.
On solving the above differential equation, one can find
\begin{equation}\label{2b7}
e^{\lambda(r)}=1+c_1e^{\xi(r)}\xi^{'2}(r),
\end{equation}
where $c_1$ is the constant of integration. Using the above expression \eqref{2b7}, we shall obtain wormhole shape functions for different redshift functions in section-\ref{sec5}.

\section{Dark matter halos}\label{sec4}
Several inquiries are highlighted herein in the DM halos context, particularly those involving galactic wormholes, primarily deduced from rotation curves \cite{Islam1,Islam3}. It is essential to acknowledge that the postulation of ``dark matter" arose from the review of Oort constants \cite{Oort4} and investigations into the masses of galaxy clusters \cite{Zwicky1,Zwicky2}, aiming to explain the observed flat rotation curves. To address this phenomenon, the Navarro-Frenk-White density profile function \cite{Frenk2,Frenk3} and, alternatively, the Einasto profile was suggested, with the latter demonstrating improved alignment with specific dark matter halo simulations \cite{Frenk4,Frenk5}. Readers interested in galactic wormholes are encouraged to refer to some of the recent relevant literature \cite{Rizwan1,Rizwan2,Rizwan4}. Generally, the wormhole throat is surrounded by varying types of DM halos. The observed deviation from expected energy conditions substantiates the presence of these dark halos. This study explores the representations associated with different kinds of dark matter halos, incorporating wormhole solutions within various DM halo profiles in this section.
\subsection{Universal Rotation Curve (URC) dark matter profile}
For the first case, we consider the URC DM halo profile, which is defined by \cite{Burkertt}
\begin{equation}
\rho(r)=\frac{\rho_s\,r_s^3}{(r+r_s)(r^2+r_s^2)},
\end{equation}
where $r_s$ and $\rho_s$ denote the core radius and central density, respectively. The subsequent aspect under consideration is the rotation curve, a parameter for which the analysis is widely regarded as a significant pillar supporting the presence of dark matter in galaxies. Persic et al. \cite{Stel}, in their investigation utilizing $H_\alpha$ data and incorporating certain radio rotation curves, have observed an extensive array of rotation curves. Their findings suggest that rotation curves can be effectively accommodated for various luminosities and across various galaxy types, encompassing spirals, low-surface-brightness ellipticals, and dwarf-irregular galaxies. Consequently, they introduced the term ``universal rotation curve" to replace the conventional term ``rotation curves," reflecting the broad applicability of the phenomenon.
\subsection{Navarro-Frenk-White (NFW) dark matter profile}
An approximate analytical formulation of the NFW density profile is established by drawing upon the Cold Dark Matter ($\Lambda$CDM) theory and numerical simulations \cite{Carlberg,Frenk2,Frenk3}. In the context of galaxies and clusters, the dark matter halo can be characterized through the NFW density profile, expressed as
\begin{equation}
\rho(r)=\frac{\rho_s}{(r/r_s)(1+r/r_s)^2},
\end{equation}
where $\rho_s$ denotes the dark matter density during the collapse of the dark matter halo, and $r_s$ represents the scale radius. It is widely recognized that the NFW density profile encompasses a diverse range of dark matter models characterized by minimal collision effects between dark matter particles.
\subsection{Scalar Field Dark Matter (SFDM) profile}
The SFDM model contains two distinct density profiles, namely the Bose-Einstein Condensate (BEC) profile and the finite BEC profile \cite{Matos1,Matos2}. For the sake of simplicity, we concentrate on the BEC profile. This particular profile aligns with the static solution of the Klein-Gordon equation and involves a quadratic potential governing the scalar field. The corresponding SFDM profile \cite{Hou1,Hou2} of DM halo is given by
\begin{equation}
\rho(r)=\frac{\rho_s \sin (\pi r/r_s)}{\pi r/r_s},
\end{equation}
where $r_s$ denotes the radius at which the pressure and density are zero, and $\rho_s$ is the central density.

\section{Wormhole solutions for different redshift functions}\label{sec5}
In this section, we will explore wormhole solutions with non-zero tidal force solutions, where $\phi(r)\neq 0$. A non-constant redshift function can improve the stability of a wormhole and enhance its traversability. By appropriately choosing the redshift function, one can control the gravitational tidal forces experienced by travelers. For instance, a well-chosen redshift function can ensure that these tidal forces remain weak enough for safe passage through the wormhole, which is essential for the traversability condition. However, many researchers have focused on the zero tidal force (ZTF) model, where $\phi(r)$ is kept constant, to simplify the calculations of the energy conditions. Maintaining a constant redshift may not be realistic from a physics perspective. Therefore, removing this assumption and allowing for a non-constant redshift function is crucial. In \cite{Liempi1}, the authors confirm that the ZTF condition does not enable wormholes to exist with isotropic pressure. The physical properties of traversable wormholes under a phantom energy state were investigated in \cite{Liempi2}, and it was discussed that a non-constant redshift function could help minimize the amount of exotic matter needed to sustain the wormhole. Additionally, the impact of redshift functions on the WEC within the framework of $f(R)$ gravity has been discussed in \cite{Liempi3}. Thus, a non-constant redshift function offers numerous theoretical advantages in wormhole geometry, thereby improving their feasibility and versatility as solutions within GR and beyond.\\
Motivated by the above lines, we shall consider two different non-constant redshift functions to study the properties of wormhole solutions. Now, to compute the shape functions, we compare the coefficients of the metrics \eqref{2b1} and \eqref{2c1}, and we obtain
\begin{equation}\label{41}
  \xi(r)=2\phi(r) \,\,\,\,\,\,\,\,\,\text{and}\,\,\,\,\,\,\,\,e^{\lambda(r)}=\left(1-\frac{b(r)}{r}\right)^{-1}.  
\end{equation}
Now, with the above relations, we could able to obtain the shape function from Eq. \eqref{2b7} under different redshift functions.
\subsection{$\phi(r)=-\frac{k}{r}$}
\begin{figure}[h]
    \centering
    \includegraphics[width=7.5cm,height=5cm]{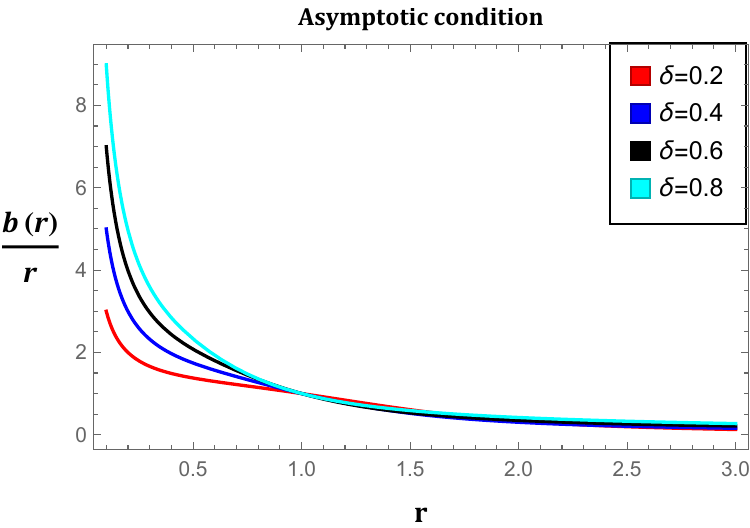}
    \caption{Asymptotically flatness condition of the wormhole against the radial distance $r$ for the redshift $\phi(r)=-\frac{k}{r}$. We use $k=0.2$ and $r_0=1$.}
    \label{fig:1}
\end{figure}

\begin{figure}[h]
    \centering
    \includegraphics[width=7.5cm,height=5cm]{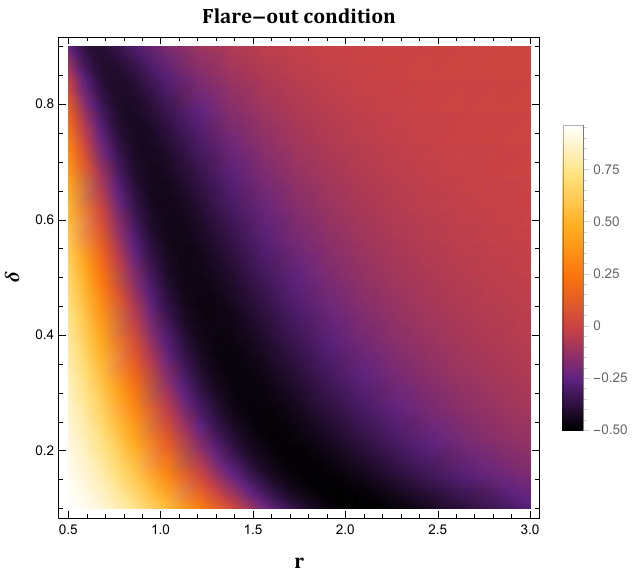}
    \caption{Flare-out condition of the wormhole against the radial distance $r$ for the redshift $\phi(r)=-\frac{k}{r}$. We use $k=0.2$ and $r_0=1$.}
    \label{fig:2}
\end{figure}

For the first case, we consider the redshift function defined by
\begin{equation}\label{4a1}
\phi(r)=-\frac{k}{r},
\end{equation}
where $k$ is any positive constant. Note that this form of redshift function satisfies the asymptotic flatness condition, i.e., $\phi(r)\rightarrow 0$ as $r\rightarrow \infty$. Kar and Sahdev \cite{Sahdev} proposed this specific redshift function and investigated the nature of matter (WEC) and the embedding of the space-like slices. They also discussed the human traversability of these space times under this redshift function. Later, L. A. Anchordoqui et al. \cite{Anchordoqui} discussed evolving wormholes with this specific redshift function. They studied some issues related to the WEC violation and human traversability in these time-dependent geometries. Additionally, with this specific redshift function, wormhole solutions have been investigated in different modified theories of gravity, including $f(R)$ gravity \cite{Fayyaz1}, $f(R,\Phi)$ gravity \cite{Fayyaz2}, and $f(Q)$ gravity \cite{Mustafa}.\\
Using Eqs. \eqref{2b7}, \eqref{41} and \eqref{4a1}, one can obtain the shape function
\begin{equation}\label{4a2}
b(r)=r-\frac{r^5}{r^4+4c_1k^2 e^{-2k/r}}.
\end{equation}
Note that when we impose the throat condition, i.e., $b(r_0)=r_0$, we get the trivial solution $r_0=0$. Thus, to avoid this issue, we add a free parameter $\delta$, and hence the shape function \eqref{4a2} becomes 
\begin{equation}\label{4a3}
b(r)=r-\frac{r^5}{r^4+4c_1k^2 e^{-2k/r}}+\delta.
\end{equation}
Now, to find the integrating constant $c_1$, we use the throat condition and obtain
\begin{equation}\label{4a4}
c_1=\frac{r_0^4(r_0-\delta)}{4\delta k^2 e^{-2k/r_0}}.
\end{equation}
Substituting the value of $c_1$ in Eq. \eqref{4a3}, one can find the final version of the shape function
\begin{equation}\label{4a5}
b(r)=r-\frac{\delta r^5}{\delta r^4+r_0^4(r_0-\delta) e^{-2k(\frac{1}{r}-\frac{1}{r_0})}}+\delta,
\end{equation}
where $0<\delta<r_0$.
Hence, in this case, the Morris-Throne wormhole metric \eqref{2c1} can be read as
\begin{multline}\label{4a6}
ds^2=-e^{-\frac{2k}{r}}dt^2+\left(\frac{\delta r^4}{\delta r^4+r_0^4(r_0-\delta) e^{-2k(\frac{1}{r}-\frac{1}{r_0})}}-\frac{\delta}{r}\right)^{-1}\\
\times dr^2+r^2d\Omega^2_{D-2}.
\end{multline}
Clearly, in the limit $r\rightarrow \infty$, we obtain
$$\lim \limits_{r\rightarrow \infty}\frac{b(r)}{r}\rightarrow 0,$$
which confirms that the asymptotically flatness condition is satisfied. Moreover, the flare-out condition is also satisfied at the throat for $0<\delta<r_0$ as well as for the appropriate choice of the other free parameters. The graphical behavior of the shape functions can be found in Figs. \ref{fig:1} and \ref{fig:2}.\\
For the redshift function \eqref{4a1}, the radial and tangential pressures can be read as
\begin{multline}\label{4a7}
p_r=\frac{1}{8 \pi}\left[\frac{1}{r^7}\left(-\delta  r \mathcal{L}_2+\alpha  \delta ^2 (r-4 k)+\alpha  r^3-r^5\right) \right.\\\left.
+\alpha  \delta ^2 r^3 (r-4 k) \mathcal{L}_1^2+\frac{\delta \mathcal{L}_1}{r^2}\left(2 k \left(4 \alpha  \delta +r^3+2 \alpha  r\right)\right.\right.\\\left.\left.
+r \left(-2 \alpha  \delta +r^3-2 \alpha  r\right)\right)\right],
\end{multline}
\begin{multline}\label{4a8}
p_t=\frac{1}{8 \pi}\left[\frac{\delta }{2 r^8}\left(-8 \alpha  \delta  k^2+\mathcal{L}_1^2\left(2 \delta  r^9 \left(k^2 \left(12 \alpha  \delta+r^3 \right.\right.\right.\right.\right.\\\left.\left.\left.\left.\left.
+12 \alpha  r\right)-k r \left(26 \alpha  \delta +r^3+26 \alpha  r\right)-2 r^2 \left(-2 \alpha  \delta+r^3 \right.\right.\right.\right.\right.\\\left.\left.\left.\left.\left.
-4 \alpha  r\right)\right)\right)+\mathcal{L}_1\left(-8 \alpha  \delta  k^2 r^5-6 \alpha  \delta  r^7-8 \alpha  r^8+4 r^{10}\right) \right.\right.\\\left.\left.
-2 k^2 r^3 -8 \alpha  k^2 r+24 \alpha  k r^2-\mathcal{L}_1^3\left(4 \alpha  \delta ^2 r^{14} (r-6 k) (2 r-k)\right)\right.\right.\\\left.\left.
+3 k r^4 +32 \alpha  \delta  k r-2 \alpha  \delta  r^2-2 \alpha  r^3+r^5\right)-\mathcal{L}_3\right],
\end{multline}
where, $\mathcal{L}_1=\frac{1}{r_0^4 (r_0-\delta ) e^{2 k \left(\frac{1}{r_0}-\frac{1}{r}\right)}+\delta  r^4}$,\\
$\mathcal{L}_2=\left(2 k \left(2 \alpha +r^2\right)+r^3-2 \alpha  r\right)$,\\
and $\mathcal{L}_3=\frac{2 \alpha  \left(\delta -\mathcal{L}_1\delta  r^5+r\right)^2}{r^6}$.\\
Now, for the URC model, the NEC can be read at the throat
\begin{equation}\label{4a9}
NEC1\bigg\vert_{r=r_0}=\frac{1}{8 \pi  r_0^4}\left[\alpha -\frac{r_0^2 \left(r_s^3 \left(1-8 \pi  \rho_s r_0^2\right)+\mathcal{L}_4\right)}{(r_s+r_0) \left(r_s^2+r_0^2\right)}\right],
\end{equation}
\begin{multline}\label{4a10}
NEC2\bigg\vert_{r=r_0}=\frac{\rho_s r_s^3}{(r_s+r_0) \left(r_s^2+r_0^2\right)}+\frac{1}{16 \pi  r_0^8}\left[2 \delta ^2  \right.\\\left.
\times (k-2 r_0)\left(k \left(4 \alpha +r_0^2\right)+r_0^3-2 \alpha  r_0\right)+\delta  r_0 (5 r_0-2 k) \right.\\\left.
\left(k \left(4 \alpha +r_0^2\right)+r_0^3-2 \alpha  r_0\right)-4 \alpha  r_0^4\right],
\end{multline}
where $\mathcal{L}_4=r_s^2 r_0+r_s r_0^2+r_0^3$.\\
Again, for the NFW model, the expression for NEC can read as
 \begin{multline}\label{4a11}
NEC1=\frac{\rho_s r_s^3}{r (r_s+r)^2}+\frac{1}{8 \pi}\left[\frac{1}{r^7}\left(-\delta  r \mathcal{L}_2+\alpha  \delta ^2 (r-4 k) \right.\right.\\\left.\left.
+\alpha  r^3-r^5\right)+\mathcal{L}_1^2\alpha  \delta ^2 r^3 (r-4 k)+\frac{\delta \mathcal{L}_1}{r^2}\left(2 k \left(4 \alpha  \delta +r^3 \right.\right.\right.\\\left.\left.\left.
+2 \alpha  r\right)+r \left(-2 \alpha  \delta +r^3-2 \alpha  r\right)\right)\right].
\end{multline}
At wormhole throat $r=r_0$, the above expression reduces to
\begin{multline}\label{4a12}
NEC1\bigg\vert_{r=r_0}=\frac{1}{8 \pi  r_0^2} \left[ \frac{\alpha}{r_0^2}-\frac{1}{(r_s+r_0)^2}\left(-8 \pi  \rho_s r_s^3 r_0+r_s^2 \right.\right.\\\left.\left.
+2 r_s r_0+r_0^2\right)\right].
\end{multline}
Similarly, we could obtain the NEC at the throat for the SFDM model as
\begin{equation}\label{4a13}
NEC1\bigg\vert_{r=r_0}=\frac{\alpha +8 \rho_s r_s r_0^3 \sin \left(\frac{\pi  r_0}{r_s}\right)-r_0^2}{8 \pi  r_0^4}.
\end{equation}
\begin{figure*}[]
\centering
\includegraphics[width=5.7cm,height=5cm]{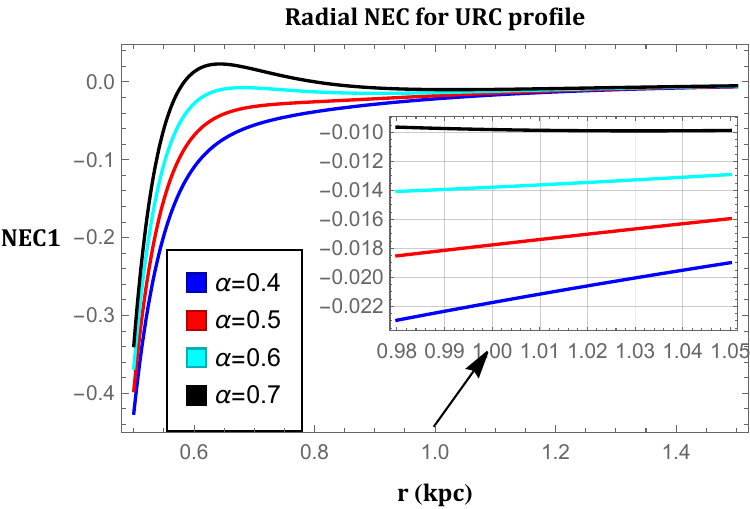}
\includegraphics[width=5.7cm,height=5cm]{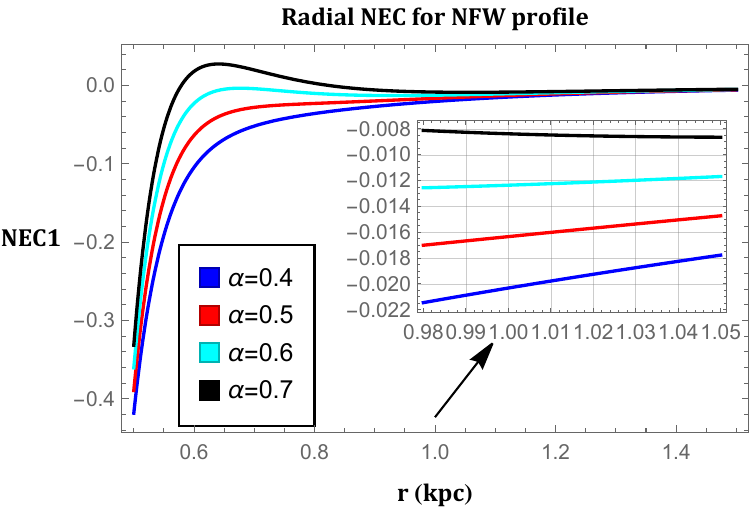}
\includegraphics[width=5.7cm,height=5cm]{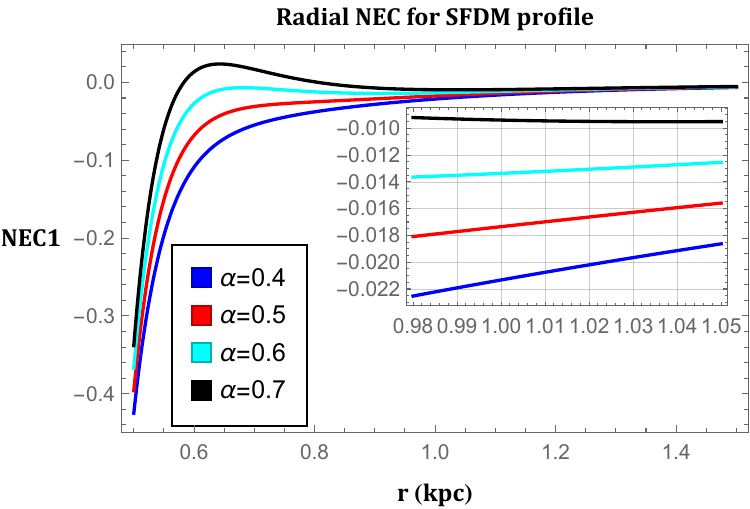}
\caption{The variation of $\rho+p_r$ against the radial distance $r$ for the redshift $\phi(r)=-\frac{k}{r}$. we use $\rho_s=0.004\, kpc^{-2}$, $r_s=2\,kpc$, $\delta=0.9$, $k=0.2$ and $r_0=1$.}
\label{fig:3}
\end{figure*}
\begin{figure*}[]
\centering
\includegraphics[width=5.7cm,height=5cm]{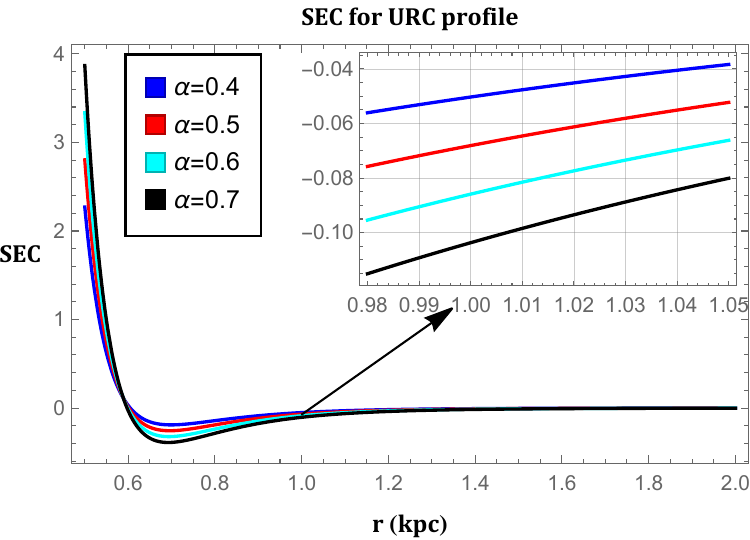}
\includegraphics[width=5.7cm,height=5cm]{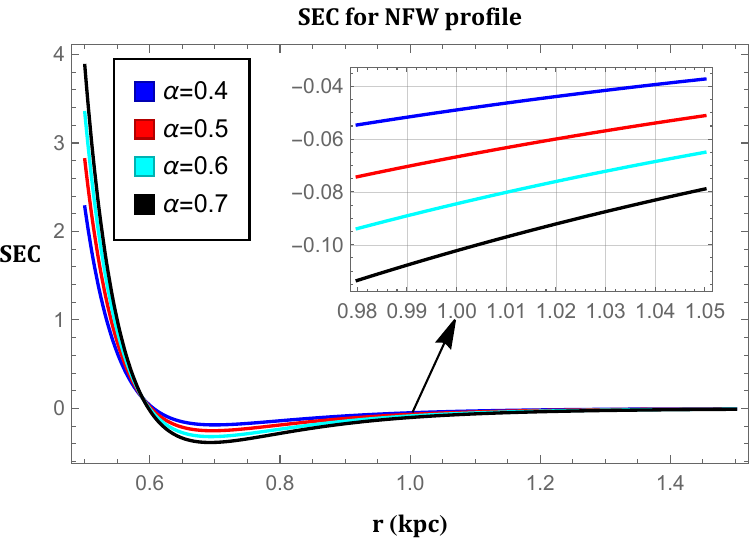}
\includegraphics[width=5.7cm,height=5cm]{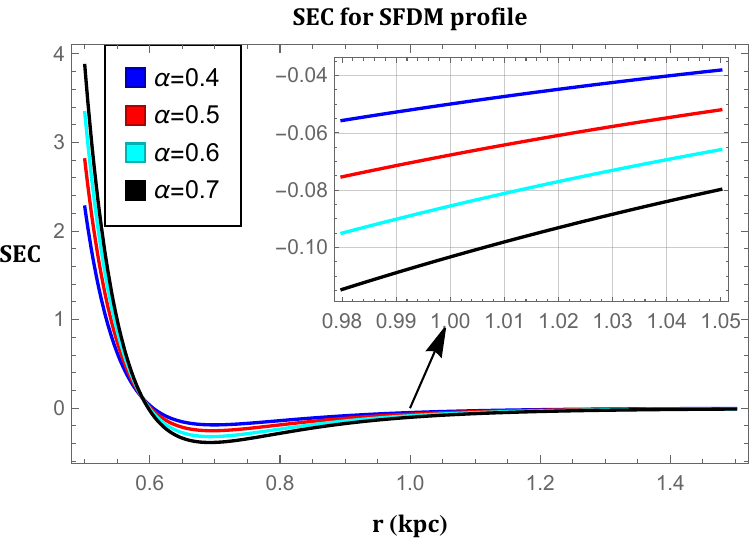}
\caption{The variation of $\rho+p_r+2p_t$ against the radial distance $r$ for the redshift $\phi(r)=-\frac{k}{r}$. we use $\rho_s=0.004\, kpc^{-2}$, $r_s=2\,kpc$, $\delta=0.9$, $k=0.2$ and $r_0=1$.}
\label{fig:4}
\end{figure*}
\begin{figure*}[]
\centering
\includegraphics[width=5.7cm,height=5cm]{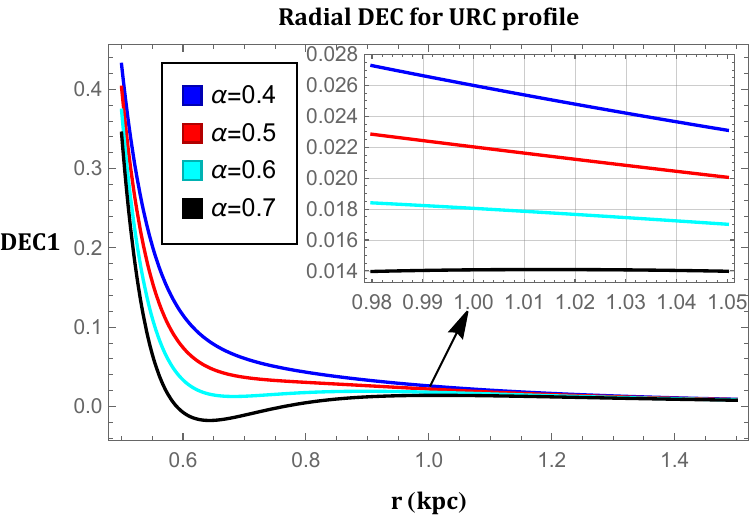}
\includegraphics[width=5.7cm,height=5cm]{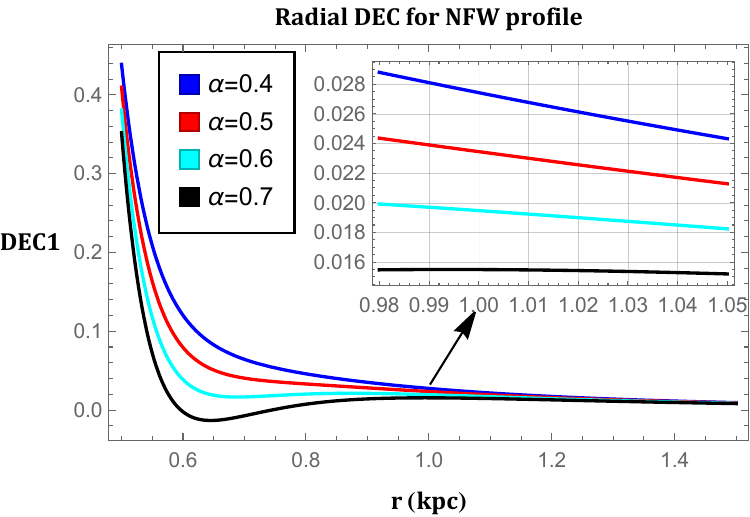}
\includegraphics[width=5.7cm,height=5cm]{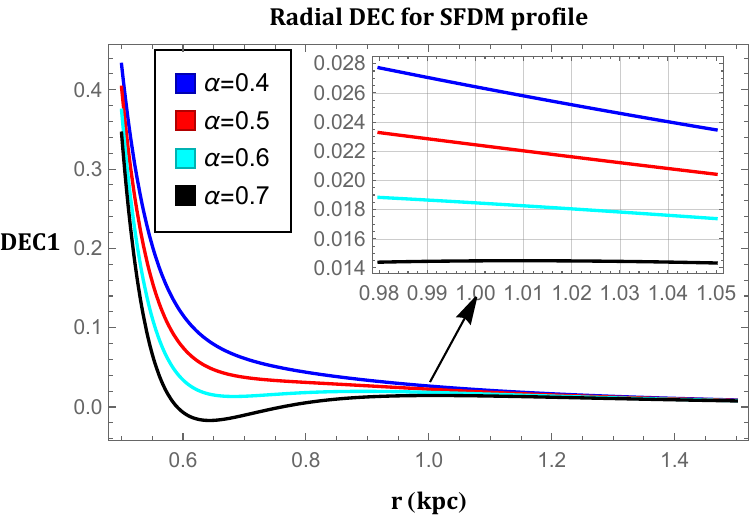}
\caption{The variation of $\rho-\abs{p_r}$ against the radial distance $r$ for the redshift $\phi(r)=-\frac{k}{r}$. we use $\rho_s=0.004\, kpc^{-2}$, $r_s=2\,kpc$, $\delta=0.9$, $k=0.2$ and $r_0=1$.}
\label{fig:5}
\end{figure*}
Now, we will discuss the energy conditions for URC, NFW, and SFDM DM halo profiles. We consider some particular choice of free parameters $\rho_s=0.004\, kpc^{-2}$, $r_s=2\,kpc$ \cite{Sarkar}, throat radius $r_0=1$, $\delta=0.9$ (as $0<\delta<r_0$) and $k=0.2$. Mathematically, one can check the RHS of Eqs. \eqref{4a9}, \eqref{4a12}, and \eqref{4a13} by simply putting the values of the parameters and taking the GB coefficient $\alpha>0$ provides a negative quantity. This confirms the violation of NEC at the wormhole throat. Graphically, we have presented the behavior of NEC for each DM profile in Fig. \ref{fig:3}. Moreover, we have studied other energy conditions, such as SEC and DEC in Figs. \ref{fig:4} and \ref{fig:5}. SEC is violated near the throat; however, far from the throat, SEC will be satisfied. Moreover, radial DEC is satisfied at the throat. But for $\alpha> > 0$, DEC will disrespect the energy conditions.

\subsection{$\phi(r)=\frac{1}{2}\log(1+\frac{\eta^2}{r^2})$}
\begin{figure}[h]
    \centering
    \includegraphics[width=7.5cm,height=5cm]{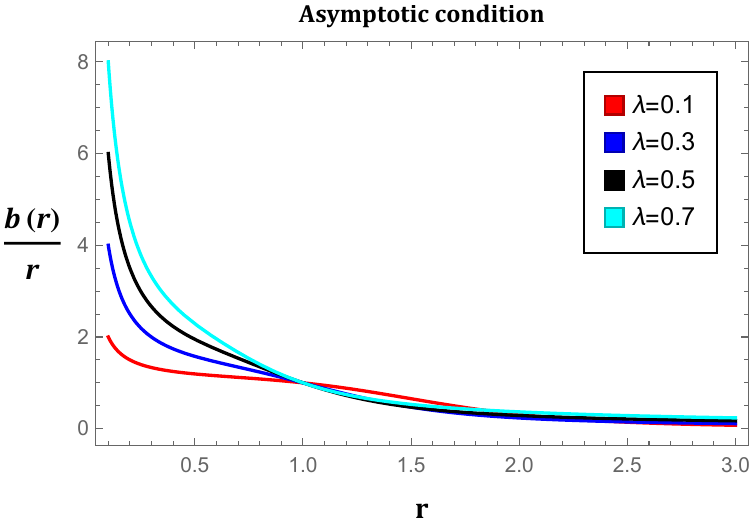}
    \caption{Asymptotically flatness condition of the wormhole against the radial distance $r$ for the redshift $\phi(r)=\frac{1}{2}\log(1+\frac{\eta^2}{r^2})$. We use $\eta=2$ and $r_0=1$.}
    \label{fig:6}
\end{figure}

\begin{figure}[h]
    \centering
    \includegraphics[width=8cm,height=5cm]{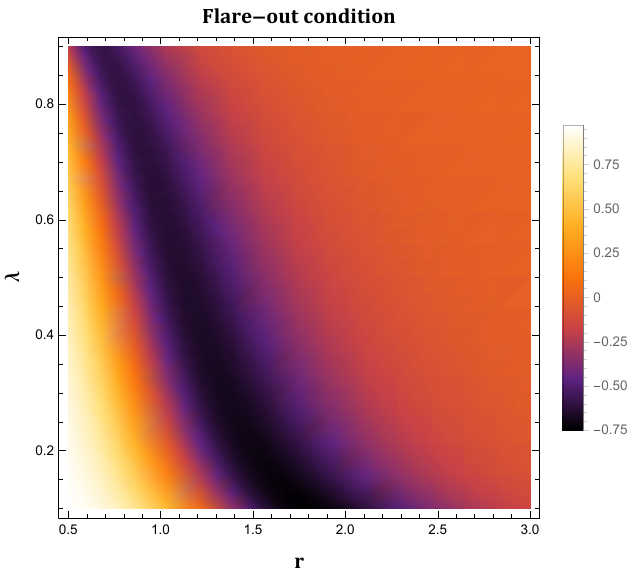}
    \caption{Flare-out condition of the wormhole against the radial distance $r$ for the redshift $\phi(r)=\frac{1}{2}\log(1+\frac{\eta^2}{r^2})$. We use $\eta=2$ and $r_0=1$.}
    \label{fig:7}
\end{figure}

\begin{figure*}[]
\centering
\includegraphics[width=5.7cm,height=5cm]{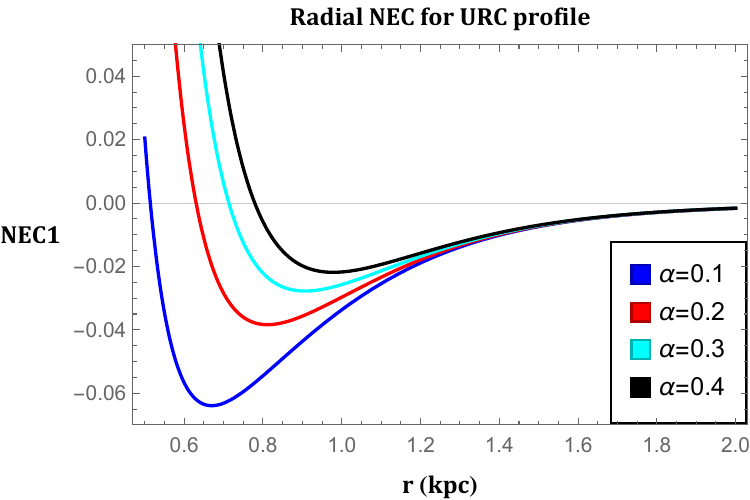}
\includegraphics[width=5.7cm,height=5cm]{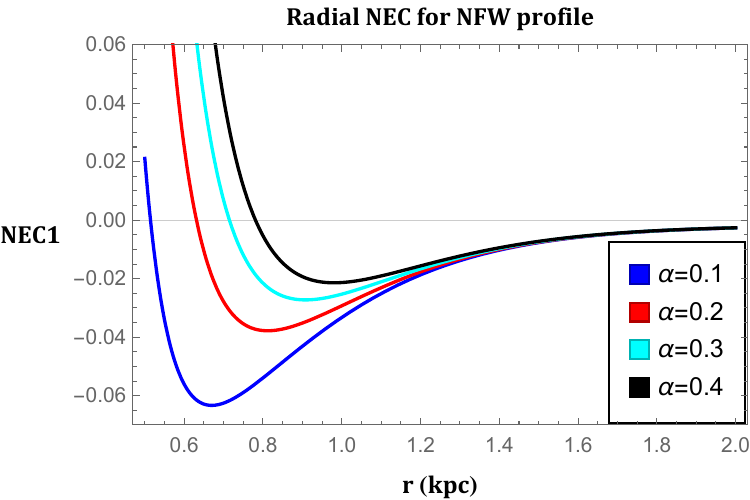}
\includegraphics[width=5.7cm,height=5cm]{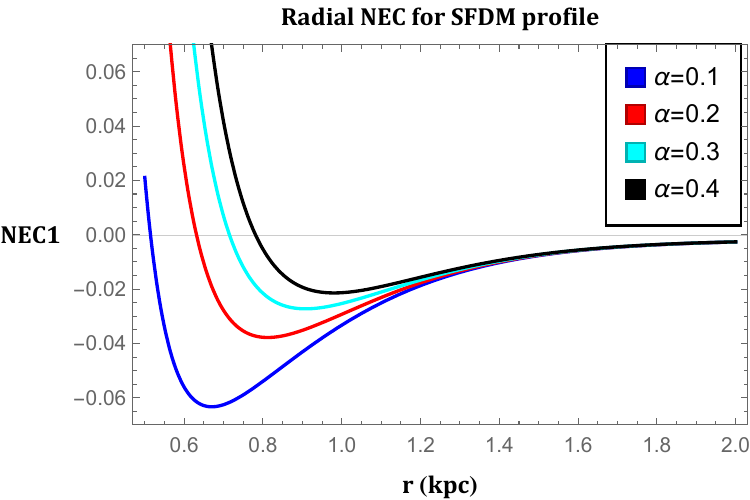}
\caption{The variation of $\rho+p_r$ against the radial distance $r$ for the redshift $\phi(r)=\frac{1}{2}\log(1+\frac{\eta^2}{r^2})$. we use $\rho_s=0.004\, kpc^{-2}$, $r_s=2\,kpc$, $\lambda=0.3$, $\eta=2$ and $r_0=1$.}
\label{fig:8}
\end{figure*}

\begin{figure*}[]
\centering
\includegraphics[width=5.7cm,height=5cm]{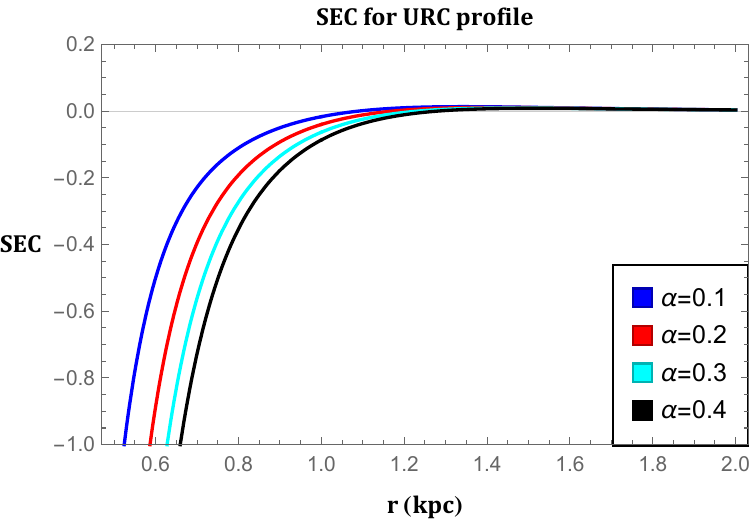}
\includegraphics[width=5.7cm,height=5cm]{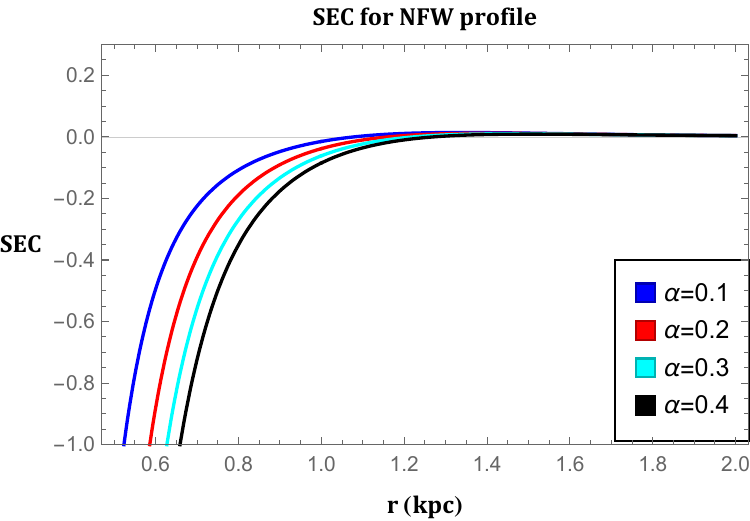}
\includegraphics[width=5.7cm,height=5cm]{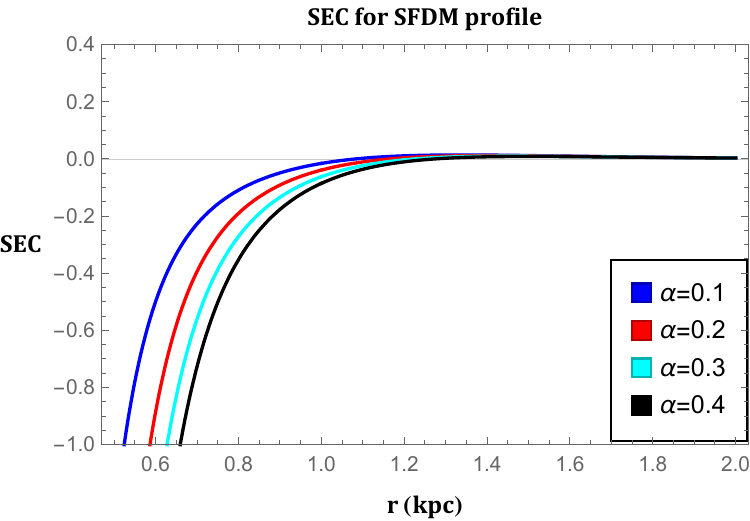}
\caption{The variation of $\rho+p_r+2p_t$ against the radial distance $r$ for the redshift $\phi(r)=\frac{1}{2}\log(1+\frac{\eta^2}{r^2})$. we use $\rho_s=0.004\, kpc^{-2}$, $r_s=2\,kpc$, $\lambda=0.3$, $\eta=2$ and $r_0=1$.}
\label{fig:9}
\end{figure*}

\begin{figure*}[]
\centering
\includegraphics[width=5.7cm,height=5cm]{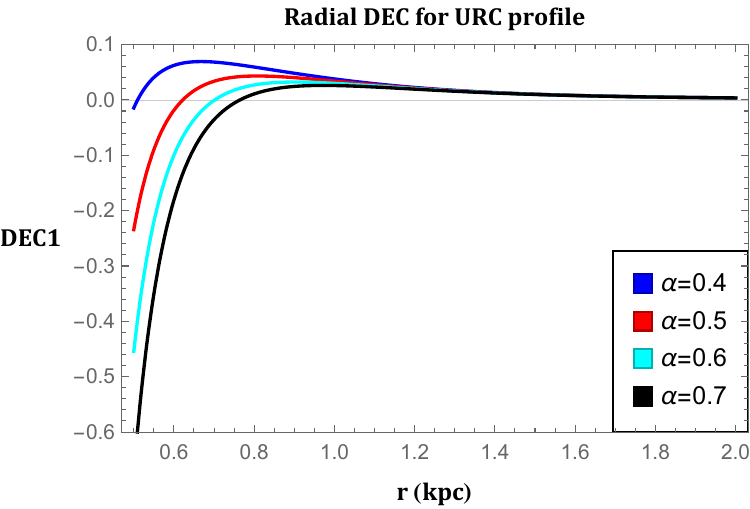}
\includegraphics[width=5.7cm,height=5cm]{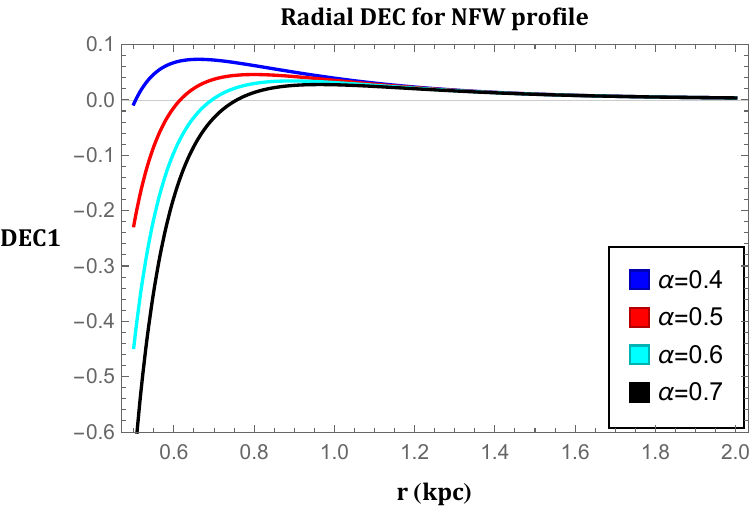}
\includegraphics[width=5.7cm,height=5cm]{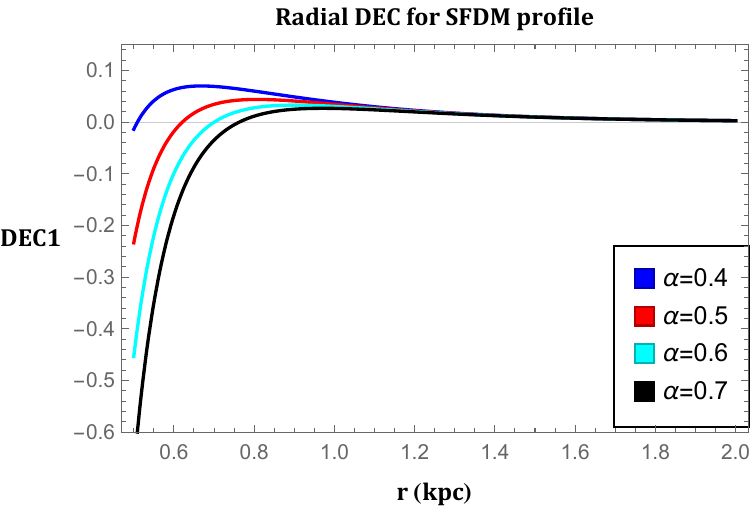}
\caption{The variation of $\rho-\abs{p_r}$ against the radial distance $r$ for the redshift $\phi(r)=\frac{1}{2}\log(1+\frac{\eta^2}{r^2})$. we use $\rho_s=0.004\, kpc^{-2}$, $r_s=2\,kpc$, $\lambda=0.3$, $\eta=2$ and $r_0=1$.}
\label{fig:10}
\end{figure*}

In this case, we consider the redshift function of the form
\begin{equation}\label{4b1}
\phi(r)=\frac{1}{2}\log(1+\frac{\eta^2}{r^2}),
\end{equation}
where $\eta$ is any positive parameter. Note that this non-constant redshift function respects the asymptotic condition of a traversable wormhole. With the above choice of redshift function, Jusufi et al. \cite{Channuie} studied traversable wormholes supported by Casimir energy in general relativity. Also, in \cite{Hassan2}, GUP-Corrected Casimir wormholes have been discussed in modified $f(Q)$ gravity with this logarithmic redshift function. Motivated by the above articles, we consider this specific choice of redshift function to check the stability of wormhole solutions in $4D$ EGB gravity.\\
Similar to the previous subsection, we could obtain the shape function by using Eqs. \eqref{2b7}, \eqref{41} and \eqref{4b1},
\begin{equation}\label{4b2}
b(r)=r-\frac{r^5\left(r^2+\eta^2\right)}{r^4\left(r^2+\eta^2\right)+4\eta^2 c_1}.
\end{equation}
It is clear that when we impose the throat condition on the above equation, we get the trivial solution $r_0=0$. Thus, we introduced a free parameter $\lambda$ to the above shape function, and hence it becomes
\begin{equation}\label{4b3}
b(r)=r-\frac{r^5\left(r^2+\eta^2\right)}{r^4\left(r^2+\eta^2\right)+4\eta^2 c_1}+\lambda.
\end{equation}
Now, we impose $b(r_0)=r_0$ to the above expression \eqref{4b3} to obtain $c_1$
\begin{equation}\label{4b4}
c_1=\frac{\mathcal{K}_1}{4\eta^2 \lambda},
\end{equation}
where $\mathcal{K}_1=r_0^4 \left(r_0^2+\eta^2\right)\left(r_0-\lambda\right)$.
Substituting the value of $c_1$ in Eq. \eqref{4b3}, we can obtain the shape function
\begin{equation}\label{4b5}
b(r)=r-\frac{r^5 \left(r^2+\eta^2\right) \lambda}{\lambda r^4 \left(r^2+\eta^2\right)+\mathcal{K}_1}+\lambda.
\end{equation}
where $0<\lambda<r_0$.
Hence, the Morris-Throne wormhole metric \eqref{2c1} reduces to
\begin{multline}\label{4b6}
ds^2=-\left(1+\frac{\eta^2}{r^2}\right)^{-1}dt^2+\\
\left(\frac{r^4 \left(r^2+\eta^2\right) \lambda}{\lambda r^4 \left(r^2+\eta^2\right)+\mathcal{K}_1}-\frac{\lambda}{r}\right)^{-1} dr^2+r^2d\Omega^2_{D-2},
\end{multline}
which is asymptotically flat space-time. In fact, one can smoothly verify that the case $0<\lambda<r_0$ gives the asymptotically flat solution. Also, we investigate the important criteria of a traversable wormhole is the flare-out condition, which is also satisfied within $0<\lambda<r_0$. In this case, we consider the throat radius $r_0=1$ and $\eta=2$. The graphical behavior of shape functions can be found in Figs. \ref{fig:6} and \ref{fig:7}.\\
For the given redshift function \eqref{4b1} and shape function \eqref{4b5}, we can compute the components of the energy-momentum tensor. In fact, the radial pressure can be read as follows
\begin{multline}\label{4b7}
p_r=\frac{1}{8 \pi  r^6}\left[\alpha  \left(\lambda -\frac{4 \lambda ^3 \left(\mathcal{K}_2-1\right)}{r^2 \left(\frac{\lambda ^2}{r^2}+1\right)}-\lambda \mathcal{K}_2 +r\right) \left(\lambda- \lambda \mathcal{K}_2 \right.\right.\\\left.\left.
+r\right)
-\frac{2 \lambda ^3 r \left(\mathcal{K}_2-1\right)}{\frac{\lambda ^2}{r^2}+1}-\left(r^3 \left(\lambda -\lambda  \mathcal{K}_2+r\right)\right)\right],
\end{multline}
where $\mathcal{K}_2=\frac{r^5 \left(\eta ^2+r^2\right)}{\lambda  r^4 \left(\eta ^2+r^2\right)+r_0^4 \left(\eta ^2+r_0^2\right) (r_0-\lambda )}$.\\
On the other hand, tangential pressure in this case can be read as
\begin{multline}\label{4b8}
p_t=\frac{1}{8 \pi } \left[\frac{1}{r} \left(\lambda  \left(\mathcal{K}_2-1\right) 
\left(\frac{\lambda ^2 \left(\mathcal{K}_3+1\right)}{2 r^4 \left(\frac{\lambda ^2}{r^2}+1\right) \left(1-\mathcal{K}_2\right)} \right.\right.\right.\\ \left.\left.\left.
\times \left(\frac{12 \alpha  \lambda }{r^3}+\frac{4 \alpha }{r^2}-\frac{12 \alpha  \lambda \mathcal{K}_2}{r^3}+1\right)+\frac{\left(2 \lambda ^4+3 \lambda ^2 r^2\right)}{r^2 \left(\lambda ^2+r^2\right)^2} \right.\right.\right.\\\left.\left.\left.
\times \left(\frac{4 \alpha  \left(\lambda -\lambda \mathcal{K}_2+r\right)}{r^3}+1\right)
+\frac{1}{r}\left(\left(-\frac{\lambda ^2}{r^3+\lambda ^2 r} \right.\right.\right.\right.\right.\\\left.\left.\left.\left.\left.
+\frac{\mathcal{K}_3+1}{2 r \left(\mathcal{K}_2-1\right)}\right) \left(1-\frac{2 \alpha  \left(\lambda -\lambda  \mathcal{K}_2+r\right)}{r^3}\right)\right)\right)\right)  \right.\\\left.
-\frac{2 \alpha  \left(\lambda -\lambda  \mathcal{K}_2+r\right)^2}{r^6}\right],
\end{multline}  

where $\mathcal{K}_3=\frac{2 r^5 r_0^4 \left(2 \eta ^2+3 r^2\right) \left(\eta ^2+r_0^2\right) (r_0-\lambda )}{\left(\lambda  r^4 \left(\eta ^2+r^2\right)-\eta ^2 \lambda  r_0^4+\eta ^2 r_0^5-\lambda  r_0^6+r_0^7\right)^2}$.\\
Now, with the above pressure components, we can examine the energy conditions for the URC, NFW, and SFDM dark matter halo profiles and try to generate plots to evaluate the validity of the energy conditions.\\
The NEC at wormhole throat ($r=r_0$) for each DM halo profile has been obtained and shown in Eq. \eqref{4b9}.
\begin{widetext}
\begin{equation}\label{4b9}
NEC1\bigg\vert_{r=r_0}=
     \begin{cases}
      \frac{1}{8 \pi  r_0^4}\left[\alpha -\frac{r_0^2 \left(r_s^3 \left(1-8 \pi  \rho_s r_0^2\right)+r_s^2 r_0+r_s r_0^2+r_0^3\right)}{(r_s+r_0) \left(r_s^2+r_0^2\right)}\right],  & \text{(URC profiles)}\\
      \\
      \frac{\alpha  (r_s+r_0)^2-r_0^2 \left(-8 \pi  \rho_s r_s^3 r_0+r_s^2+2 r_s r_0+r_0^2\right)}{8 \pi  r_0^4 (r_s+r_0)^2},   &  \text{(NFW profiles)}\\
      \\
      \frac{\alpha +8 \rho_s r_s r_0^3 \sin \left(\frac{\pi  r_0}{r_s}\right)-r_0^2}{8 \pi  r_0^4}, &  \text{(SFDM profiles)}
     \end{cases}
\end{equation}
\end{widetext}
Note that the above expression \eqref{4b9} is independent of the free parameters $\eta$ and $\lambda$, which confirms no influence of those parameters at the throat. However, outside the throat, we can see their influences. Also, in this case, we consider the appropriate choice of free parameters $\lambda=0.3$, $\eta=2$, $r_0=1$ along with the same values of $\rho_s$ and $r_s$, which we choose in the previous subsection. Numerically, one can check by putting the above values along with $\alpha> 0$, which gives a negative quantity of NEC. Graphically, we have presented the behavior of NEC in Fig. \ref{fig:8}. Moreover, we studied SEC in Fig. \ref{fig:9} and found that SEC is violated near the throat for $\alpha > 0$. We also noticed that as we increase the values of $\alpha$, the contribution of violation becomes more. Further, we checked DEC and observed that DEC is satisfied in the vicinity of the throat (see Fig. \ref{fig:10}).\\
A detailed summary of the energy conditions for URC, NFW, and SFDM galactic DM profiles against both redshift functions have been calculated and shown in Table- \ref{Table0}.

\begin{table*}[t]
    \centering
\begin{tabular}{ p{3cm} p{6.5cm} p{6.5cm}}
 \hline
 \multicolumn{3}{|c|}{The behavior of the energy conditions around the throat} \\
 \hline
 Physical expressions & $\phi(r)=-\frac{k}{r}$ with $\rho_s=0.004\, kpc^{-2}$, $r_s=2\,kpc$, $\delta=0.9$, $k=0.2$ and $r_0=1$ & $\phi(r)=\frac{1}{2}\log(1+\frac{\eta^2}{r^2})$ with $\rho_s=0.004\, kpc^{-2}$, $r_s=2\,kpc$, $\lambda=0.3$, $\eta=2$ and $r_0=1$\\
 \hline
\multicolumn{3}{|c|} {Energy conditions for URC profile}\\
\hline
$\rho$ & $\rho>0$ & $\rho>0$\\
\hline
$\rho+p_r$ & $\rho+p_r<0$ for $\alpha>0$ & $\rho+p_r<0$ for $\alpha>0$\\
\hline
$\rho+p_t$ & $\rho+p_t<0$ for $\alpha>0$ & $\rho+p_t<0$ for $\alpha>0$\\
\hline
$\rho+p_r+2p_t$ & $\rho+p_r+2p_t<0$ for $\alpha>0$ & $\rho+p_r+2p_t<0$ for $\alpha>0$\\
\hline
$\rho-\abs{p_r}$ & $\rho-\abs{p_r}>0$ for $\alpha>0$ & $\rho-\abs{p_r}>0$ for $\alpha>0$\\
\hline
$\rho-\abs{p_t}$ & $\rho-\abs{p_t}>0$ for $\alpha>0$ & $\rho-\abs{p_t}<0$ for $\alpha>0$\\
\hline
\multicolumn{3}{|c|} {Energy conditions for NFW profile}\\
\hline
$\rho$ & $\rho>0$ & $\rho>0$\\
\hline
$\rho+p_r$ & $\rho+p_r<0$ for $\alpha>0$ & $\rho+p_r<0$ for $\alpha>0$\\
\hline
$\rho+p_t$ & $\rho+p_t<0$ for $\alpha>0$ & $\rho+p_t<0$ for $\alpha>0$\\
\hline
$\rho+p_r+2p_t$ & $\rho+p_r+2p_t<0$ for $\alpha>0$ & $\rho+p_r+2p_t<0$ for $\alpha>0$\\
\hline
$\rho-\abs{p_r}$ & $\rho-\abs{p_r}>0$ for $\alpha>0$ & $\rho-\abs{p_r}>0$ for $\alpha>0$\\
\hline
$\rho-\abs{p_t}$ & $\rho-\abs{p_t}>0$ for $\alpha>0$ & $\rho-\abs{p_t}>0$ for $\alpha>0$\\
\hline
\multicolumn{3}{|c|} {Energy conditions for SFDM profile}\\
\hline
$\rho$ & $\rho>0$ & $\rho>0$\\
\hline
$\rho+p_r$ & $\rho+p_r<0$ for $\alpha>0$ & $\rho+p_r<0$ for $\alpha>0$\\
\hline
$\rho+p_t$ & $\rho+p_t<0$ for $\alpha>0$ & $\rho+p_t<0$ for $\alpha>0$\\
\hline
$\rho+p_r+2p_t$ & $\rho+p_r+2p_t<0$ for $\alpha>0$ & $\rho+p_r+2p_t<0$ for $\alpha>0$\\
\hline
$\rho-\abs{p_r}$ & $\rho-\abs{p_r}>0$ for $\alpha>0$ & $\rho-\abs{p_r}>0$ for $\alpha>0$\\
\hline
$\rho-\abs{p_t}$ & $\rho-\abs{p_t}>0$ for $\alpha>0$ & $\rho-\abs{p_t}>0$ for $\alpha>0$\\
\hline
\end{tabular}
\caption{Outlook of the energy conditions}
\label{Table0}
\end{table*}

\section{Some physical features of wormhole solutions}\label{sec6}
In this particular section, we study some analysis of the obtained wormhole solutions such as the complexity factor, active gravitational mass, total gravitational energy, and the embedding diagrams of the wormholes.
\subsection{Complexity factor}
A recent contribution in Ref. \cite{L. Herrera} proposed a novel complexity definition for self-gravitating fluid distributions. This definition centers on the intuitive concept that the gravitational system's least complexity should be represented by a homogeneous energy density distribution accompanied by isotropic pressure. The work in \cite{L. Herrera} reveals that in static spherically symmetric space-times, an associated scalar emerges from the orthogonal splitting of the Riemann tensor \cite{Gomez1,Gomez2}. This scalar, denoted as $\Upsilon_{TF}$, summarizes the core aspects of complexity, and its expression can be read as
\begin{equation}\label{6a1}
\Upsilon_{TF}=8\pi \Delta  -\dfrac{4 \pi}{r^{3}}\int _{0}^{r}\Tilde{r}^{3}\rho'(r) d\Tilde{r}\,,
\end{equation}
where $\Delta=p_r-p_t$. Moreover, the above expression \eqref{6a1} allows us to write the Tolman mass as
\begin{equation}
m_T=(m_T)_\Sigma \left(\frac{r}{r_\Sigma}\right)^3+r^3 \int_0^{r_\Sigma}\frac{e^{(\xi+\lambda)/2}}{\Tilde{r}}\Upsilon_{TF} d\Tilde{r},
\end{equation}
and this serves as a vital justification for defining the complexity factor using the aforementioned scalar, as it effectively incorporates all modifications arising from both energy density inhomogeneity and pressure anisotropy on the active gravitational mass. Notably, the condition of vanishing complexity $(\Upsilon_{TF}=0)$ can be fulfilled not solely in the simplest scenario of isotropic and homogeneous systems but also across all cases where
\begin{equation}
\Delta=\dfrac{1}{2r^{3}}\int _{0}^{r}\Tilde{r}^{3}\rho'(r) d\Tilde{r}.
\end{equation}
In this context, fulfilling the vanishing complexity condition results in a non-local equation of state, providing an additional condition to close the system of Einstein's field equations. This application has been presented in some recent papers, as seen in Refs. \cite{Casadio1,Casadio2,Casadio3}.\\
In this paper, we are interested in studying the complexity factor for the wormhole solutions under galactic regions. It is known that wormholes are defined for $r_0\leq r <\infty$, in that case the complexity factor \eqref{6a1} should be modified and can be redefined as
\begin{equation}\label{6a2}
\Upsilon_{TF}=8\pi \Delta  -\dfrac{4 \pi}{r^{3}}\int _{r_0}^{r}\Tilde{r}^{3}\rho'(r) d\Tilde{r}.
\end{equation}
Note that the standard definition requires $r_0 = 0$, but we must discard this case to ensure a finite size of the wormhole throat.\\
\begin{figure*}[]
\centering
\includegraphics[width=5.7cm,height=5cm]{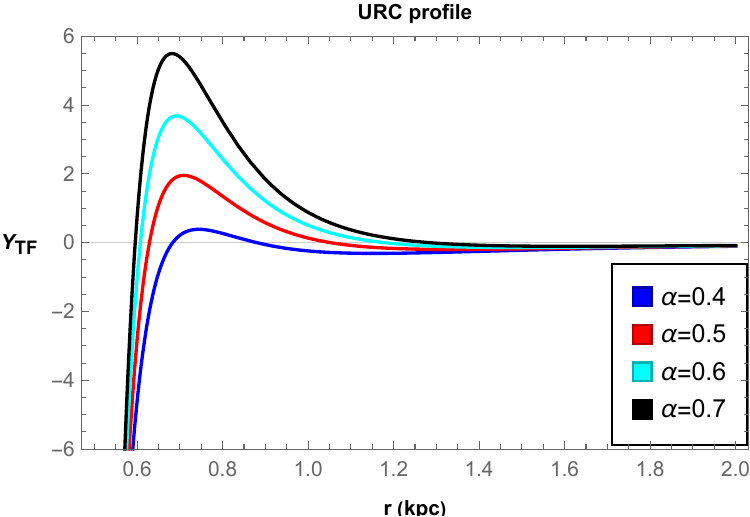}
\includegraphics[width=5.7cm,height=5cm]{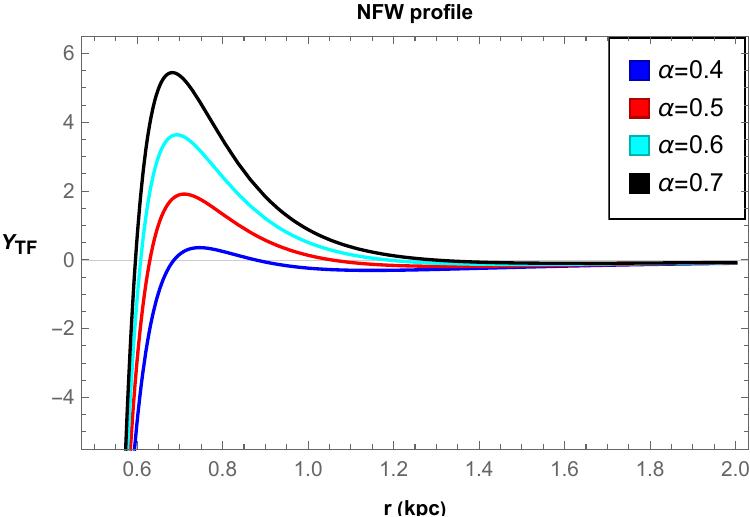}
\includegraphics[width=5.7cm,height=5cm]{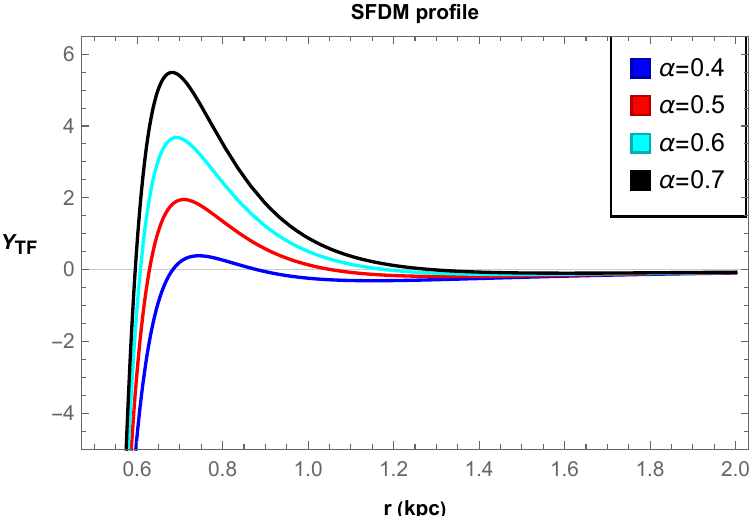}
\caption{The dynamics of $\Upsilon_{TF}$ against the radial distance $r$ for the redshift $\phi=\frac{-k}{r}$.  we use $\rho_s=0.004\, kpc^{-2}$, $r_s=2\,kpc$, $\delta=0.9$, $k=0.2$ and $r_0=1$.}
\label{fig:11}
\end{figure*}
\begin{figure*}[]
\centering
\includegraphics[width=5.7cm,height=5cm]{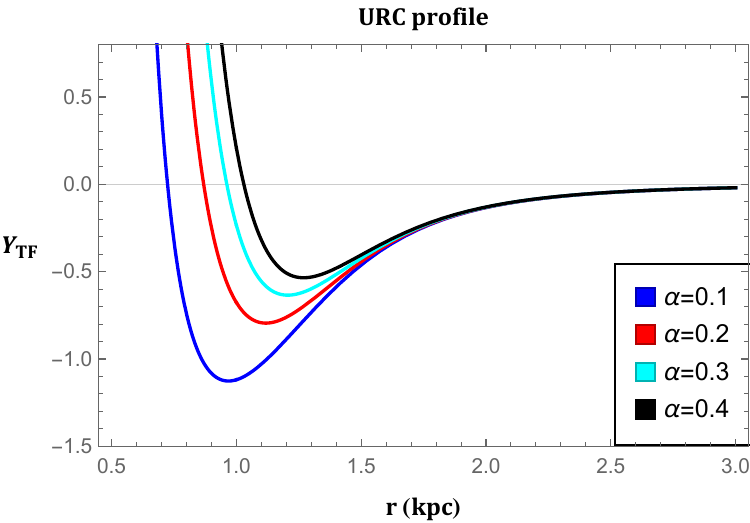}
\includegraphics[width=5.7cm,height=5cm]{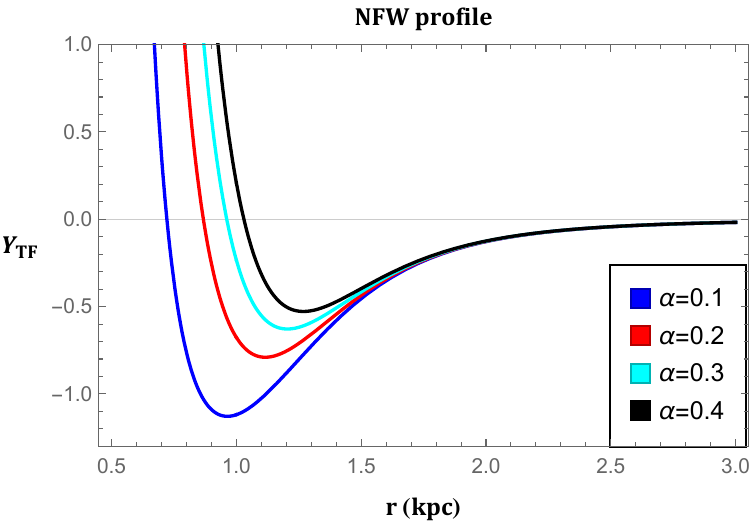}
\includegraphics[width=5.7cm,height=5cm]{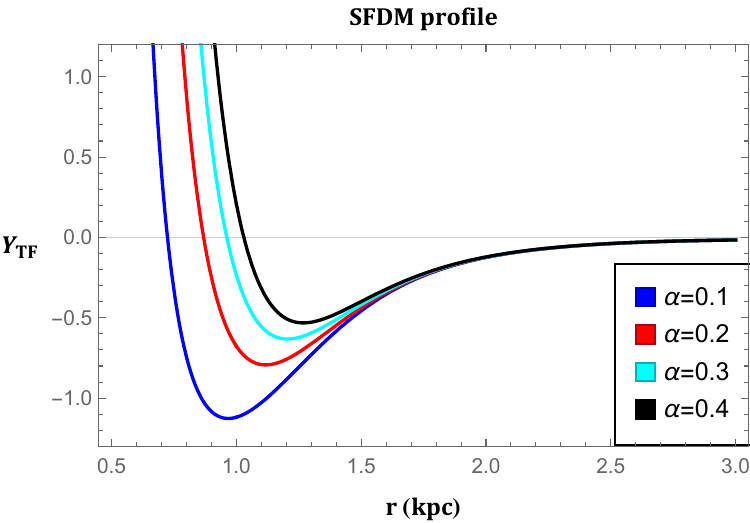}
\caption{The dynamics of $\Upsilon_{TF}$ against the radial distance $r$ for the redshift $\phi(r)=\frac{1}{2}\log(1+\frac{\eta^2}{r^2})$. we use $\rho_s=0.004\, kpc^{-2}$, $r_s=2\,kpc$, $\lambda=0.3$, $\eta=2$ and $r_0=1$.}
\label{fig:12}
\end{figure*}
Now, considering the relevant equations, we have studied the behavior of the complexity factor (using the above Eq. \eqref{6a2}) for three galactic halos wormholes with respect to the radial coordinate shown in Figs. \ref{fig:11} and \ref{fig:12}. We noticed that the complexity factor $\Upsilon_{TF}$ tends toward zero as the radial coordinate approaches infinity $r \rightarrow \infty$ or moves away from the wormhole throat. As explained in \cite{L. Herrera}, a minimal complexity factor, corresponds to a configuration characterized by homogeneous energy density and isotropic pressure. Additionally, a complexity factor of zero suggests the presence of inhomogeneous energy density and anisotropic pressure, provided these two effects counterbalance each other in the overall complexity factor. Consequently, in the vicinity of the wormhole throat, the complexity factor exhibits a monotonically increasing direction, and as we increase the range of the radial coordinate, $\Upsilon_{TF}$ approaches zero. Moreover, we noticed that the above holds for any values of $\alpha>0$. Therefore, the complexity factor converges to zero for high radial coordinates and $\alpha>0$ in the context of galactic halo wormholes in EGB gravity. Further, in the dynamics of the complexity factor, the role of pressure isotropy emerges as more crucial compared to the homogeneity of energy density. Furthermore, one can read some recent interesting papers on this topic given in Refs. \cite{Butt1,Butt2,Butt3}.

\subsection{Active gravitational mass}
The active mass function for our wormhole within the region from the wormhole throat $r_0$ up to the radius $R$ can be read as
\begin{equation}\label{6b1}
\mathcal{M}_{\mathcal{A}}=4 \pi \int_{r_0}^{R}\rho(r) r^2 dr,
\end{equation}
where $\mathcal{M}_{\mathcal{A}}$ is the active gravitational mass. Note that the positive
nature of the active gravitational mass indicates that the models are physically acceptable. Keeping this in mind, we shall investigate three different types of DM halo models and try to find some conditions for which one can find positive mass function. \\
For the URC profile, the expression for $\mathcal{M}_{\mathcal{A}}$ can be read as
\begin{multline}\label{6b2}
\mathcal{M}_{\mathcal{A}}=4 \pi  \rho_s r_s^3 \left[\frac{1}{4} \log \left(r_s^2+r^2\right)+\frac{1}{2} \log (r_s+r)\right.\\\left.
-\frac{1}{2} \tan ^{-1}\left(\frac{r}{r_s}\right)\right]_{r_0}^{R}.
\end{multline}
It is observed from the above expression \eqref{6b2} that the active gravitational mass $\mathcal{M}_{\mathcal{A}}$ of the wormhole for URC galactic is positive under the constraint $\frac{1}{4} \log \left(r_s^2+r^2\right)+\frac{1}{2} \log (r_s+r)> \frac{1}{2} \tan ^{-1}\left(\frac{r}{r_s}\right)$.\\
Again, for the NFW profile, we can obtain the active gravitational mass
\begin{equation}\label{6b3}
\mathcal{M}_{\mathcal{A}}=4 \pi  \rho_s r_s^3 \left[\frac{r_s}{r_s+r}+\log (r_s+r)\right]_{r_0}^{R}.
\end{equation}
Note that the RHS of the Eq. \eqref{6b3} is a positive quantity for any values of the parameters.\\
At last, for the SFDM profile, $\mathcal{M}_{\mathcal{A}}$ can be read as
\begin{equation}\label{6b4}
\mathcal{M}_{\mathcal{A}}=4 \rho_s r_s \left[\frac{r_s^2 \sin \left(\frac{\pi  r}{r_s}\right)}{\pi ^2}-\frac{r_s r \cos \left(\frac{\pi  r}{r_s}\right)}{\pi }\right]_{r_0}^{R}.
\end{equation}
In this case, the active gravitational mass is positive under the constraint $\frac{r_s^2 \sin \left(\frac{\pi  r}{r_s}\right)}{\pi ^2}>\frac{r_s r \cos \left(\frac{\pi  r}{r_s}\right)}{\pi }$.\\
Thus, we can conclude that the above DM models are physically acceptable under some restrictions.

\subsection{Total gravitational energy}
\begin{figure}[h]
    \centering
    \includegraphics[width=7.5cm,height=5cm]{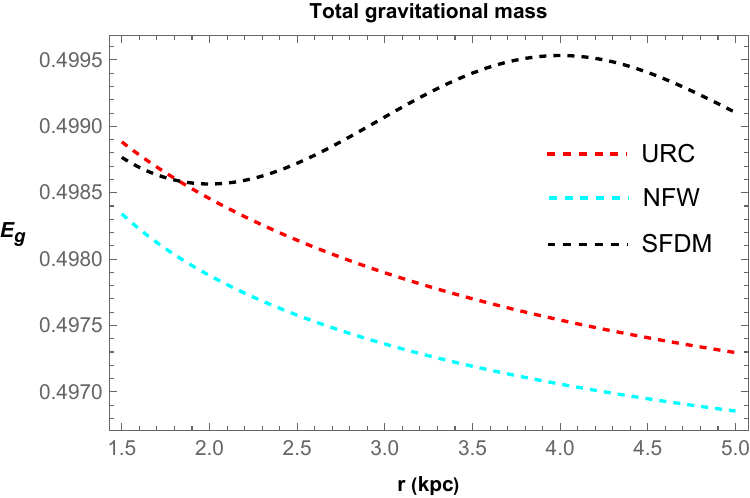}
    \caption{The variation of $\mathcal{E}_g$ under the shape function \eqref{4a5}. We use $\rho_s=0.004\, kpc^{-2}$, $r_s=2\,kpc$, $\delta=0.9$, $k=0.2$ and $r_0=1$.}
    \label{fig:13}
\end{figure}

\begin{figure}[h]
    \centering
    \includegraphics[width=7.5cm,height=5cm]{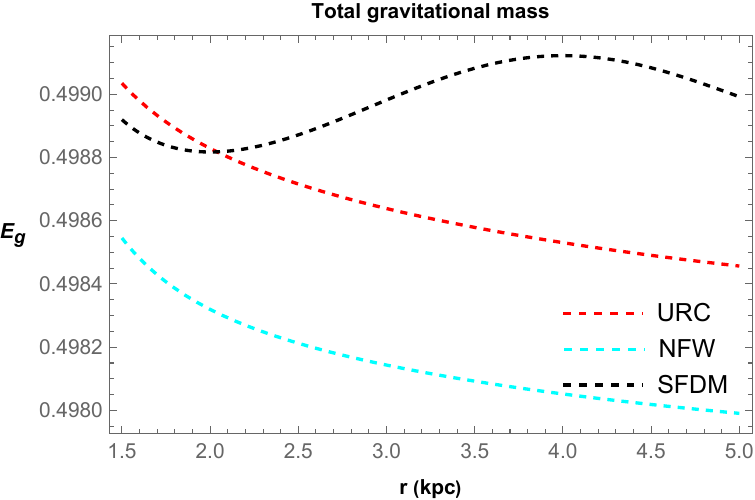}
    \caption{The variation of $\mathcal{E}_g$ under the shape function \eqref{4b5}. We use $\rho_s=0.004\, kpc^{-2}$, $r_s=2\,kpc$, $\lambda=0.3$, $\eta=2$ and $r_0=1$.}
    \label{fig:14}
\end{figure}

We have already noticed that the material forming the wormhole violates the NEC and, therefore, must be exotic rather than normal matter. The total gravitational energy of a structure composed of normal baryonic matter is negative. Therefore, it is crucial to check the nature of the gravitational energy in a wormhole background. Here, we follow the works of Lyndell-Bell et al. \cite{Katz1}, and Nandi et al. \cite{Katz2} and will try to find out the total gravitational energy of the DM galactic wormholes. The total gravitational energy $\mathcal{E}_g$ can be define as \cite{Katz2}
\begin{equation}\label{6c1}
\mathcal{E}_g=\mathcal{M}c^2-\mathcal{E}_M,
\end{equation}
where $\mathcal{M}c^2$ represents the total energy, and it can be expressed as
$$\mathcal{M}c^2=\frac{1}{2}\int_{r_0}^{r}\mathcal{T}_0^0 r^2 dr+\frac{r_0}{2},$$
where the quantity $\frac{r_0}{2}$ is linked to the effective mass \cite{Katz2}.  $\mathcal{E}_M$ is the sum of other forms of energy like kinetic energy, rest energy, internal energy, etc., defined by
$$\mathcal{E}_M=\frac{1}{2}\int_{r_0}^{r} \mathcal{T}_0^0 (g_{rr})^{\frac{1}{2}} r^2 dr,\,\,\,\text{with} \quad g_{rr}=\left(1-\frac{b(r)}{r}\right)^{-1}.$$
It can be claimed that since $(g_{rr})^{\frac{1}{2}}>1$ (by definition), then one can instantly deduce the criteria that $\mathcal{E}_g<0$ (attractive) if $\mathcal{T}_0^0>0$ and $\mathcal{E}_g>0$ (repulsive) if $\mathcal{T}_0^0<0$ \cite{A. Wheeler}. Since it is very complicated to find the exact solutions of the integral \eqref{6c1}, we solve it numerically by setting the integration range from the throat at $r_0$ to the embedded radial space of the wormhole geometry (refer to Tables \eqref{Table1} and \eqref{Table2}). As depicted in Figs. \ref{fig:13} and \ref{fig:14}, the results demonstrate that $\mathcal{E}_g>0$ signifies the repulsive behavior in the proximity of the throat. Note that we obtain $\mathcal{E}_g>0$ here despite $\mathcal{T}_0^0>0$. This is because the matter distribution supporting the wormhole structure violates the NEC. Similar behavior of $\mathcal{E}_g$ can be found in \cite{Manna1,Manna2,Manna3}. This repulsive nature of $\mathcal{E}_g$ aligns with expectations for the formation of a physically viable wormhole.
\begin{table}[t]
    \centering
\begin{tabular}{ |p{1.5cm} p{1.5cm} p{1.5cm} p{1.5cm}| }
 \hline
 \multicolumn{4}{|c|}{The values of $\mathcal{E}_g$ for different $r$.} \\
 \hline
 $r$ & URC & NFW & SFDM\\
 \hline
 1.5 & 0.498883 & 0.498341 & 0.498766\\
 2.0 & 0.498456 & 0.497876 & 0.498566\\
 2.5 & 0.498142 & 0.497577 & 0.498722\\
 3.0 & 0.497897 & 0.49736  & 0.499071\\
 3.5 & 0.497701 & 0.497193 & 0.499401\\
 4.0 & 0.497541 & 0.497058 & 0.499534\\
 4.5 & 0.497408 & 0.496948 & 0.499406\\
 5.0 & 0.497296 & 0.496855 & 0.499104\\
 \hline
\end{tabular}
\caption{Table shows the values of $\mathcal{E}_g$ for different DM profiles under the shape function \eqref{4a5}. We use $\rho_s=0.004\, kpc^{-2}$, $r_s=2\,kpc$, $\delta=0.9$, $k=0.2$ and $r_0=1$.}
\label{Table1}
\end{table}

\begin{table}[t]
    \centering
\begin{tabular}{ |p{1.5cm} p{1.5cm} p{1.5cm} p{1.5cm}| }
 \hline
 \multicolumn{4}{|c|}{The values of $\mathcal{E}_g$ for different $r$.} \\
 \hline
 $r$ & URC & NFW & SFDM\\
 \hline
 1.5 & 0.499035 & 0.498545 & 0.49892\\
 2.0 & 0.498829 & 0.49832 & 0.498818\\
 2.5 & 0.498716 & 0.498212 & 0.498872\\
 3.0 & 0.498639 & 0.498143  & 0.498982\\
 3.5 & 0.498579 & 0.498092 & 0.499083\\
 4.0 & 0.498531 & 0.498052 & 0.499122\\
 4.5 & 0.498491 & 0.498019 & 0.499084\\
 5.0 & 0.498457 & 0.497991 & 0.498992\\
 \hline
\end{tabular}
\caption{Table shows the values of $\mathcal{E}_g$ for different DM profiles under the shape function \eqref{4b5}. We use $\rho_s=0.004\, kpc^{-2}$, $r_s=2\,kpc$, $\lambda=0.3$, $\eta=2$ and $r_0=1$.}
\label{Table2}
\end{table}
\subsection{Embedding diagram}
\begin{figure}[h]
    \centering
    \includegraphics[width=7.5cm,height=5cm]{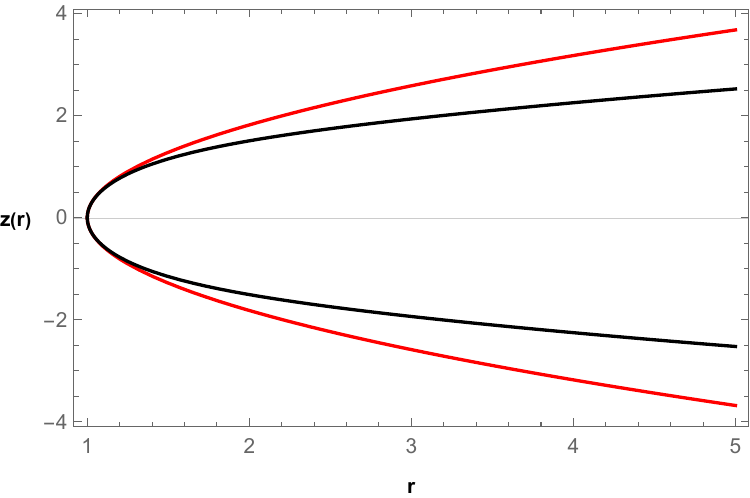}
    \caption{$2D$ embedding diagram for the shape functions \eqref{4a5} (\textit{Red}) and \eqref{4b5} (\textit{Black}). We use $\delta=0.9$, $k=0.2$, $\lambda=0.3$, $\eta=2$ and $r_0=1$.}
    \label{fig:15}
\end{figure}

\begin{figure}[h]
    \centering
    \includegraphics[width=3cm,height=5cm]{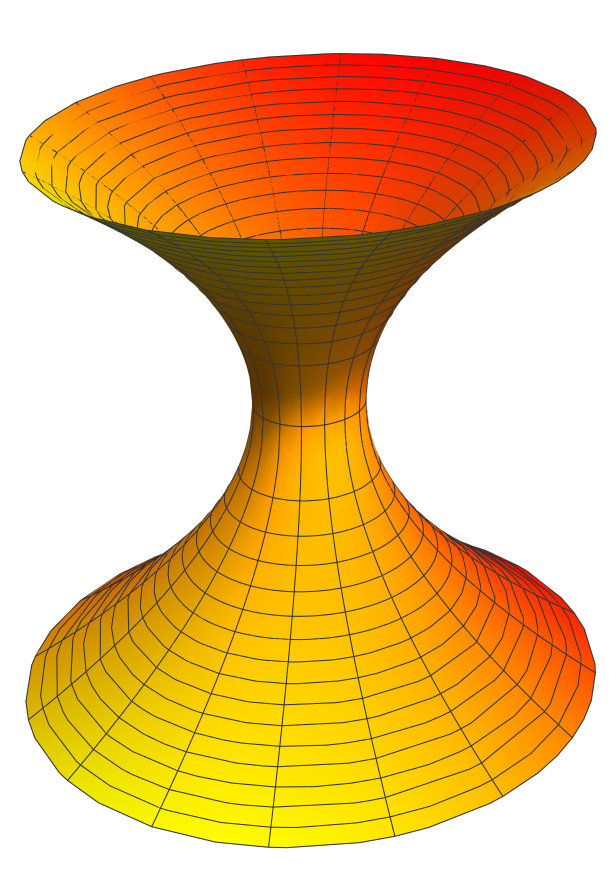}
    \includegraphics[width=5.5cm,height=5cm]{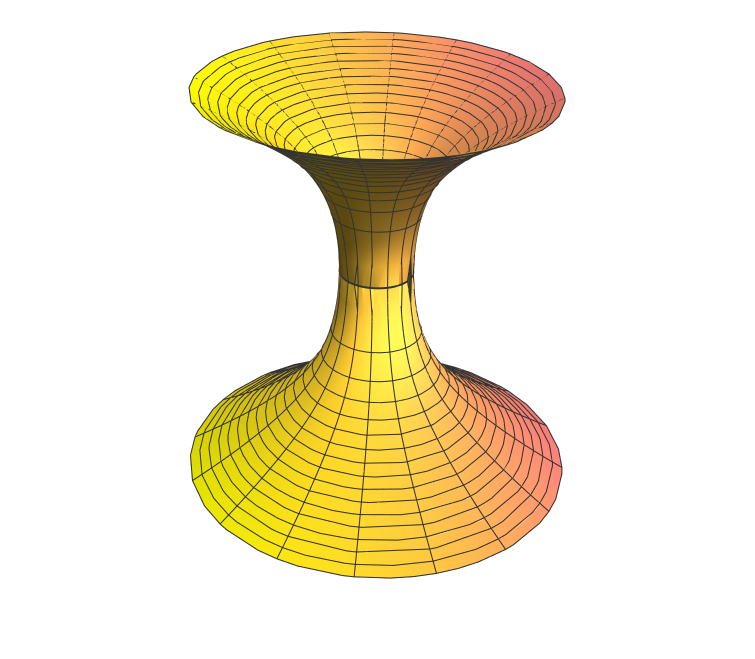}
    \caption{Full visualization of the embedding diagram for the shape functions \eqref{4a5} (\textit{left}) and \eqref{4b5} (\textit{right}). Left panel: we use $r_0=1$, $\delta=0.9$, $k=0.2$. Right panel: we use $r_0=1$, $\lambda=0.3$, $\eta=2$.}
    \label{fig:16}
\end{figure}

This subsection will discuss the embedding diagram that helps us visualize wormhole space-time. As per the framework outlined by Morris and Thorne \cite{Thorne/1988}, the embedding surface of the wormhole is represented by the function $z(r)$, which satisfies the given differential equation:
\begin{equation}\label{6d1}
\frac{dz}{dr}=\pm \frac{1}{\sqrt{\frac{r}{b(r)}-1}}.
\end{equation}
In this expression, it is observed that $\frac{dz}{dr}$ diverges at the wormhole's throat, implying that the embedding surface assumes a vertical orientation at the throat. Note that the above differential equation \eqref{6d1} provides the following relation
\begin{equation}\label{6d2}
z(r)=\pm \int_{r_0}^{r} \frac{dr}{\sqrt{\frac{r}{b(r)}-1}}.
\end{equation}
Further, the radial distance of the wormhole can be read as
\begin{equation}\label{6d3}
l(r)=\pm \int_{r_0}^{r} \frac{dr}{\sqrt{1-\frac{r}{b(r)}}}.
\end{equation}
Note that the above integral is given in Eqs. \eqref{6d2} cannot be solved analytically. Hence, we will solve it numerically by fixing some values of the free parameters and by changing the upper limit $r$. The numerical plot for embedding diagram $z(r)$ for the shape functions \eqref{4a5} and \eqref{4b5} is given in Fig. \ref{fig:15}. Moreover, one can check Fig. \ref{fig:16} for the full visualization of wormholes.


\section{Discussions and conclusions}\label{sec7}
Indeed, exploring wormhole geometry has recently sparked considerable enthusiasm among theoretical researchers. Consequently, wormholes have been identified in the galactic halo region, supported by various dark matter density profiles. This paper searches for wormhole existence in the galactic halo supported by three different DM profiles, such as URC, NFW, and SFDM models within recently proposed $4D$ EGB gravity. Also, we employed the Karmarkar condition to find the shape function for two different non-constant redshift functions. The detailed outcomes of this work are as follows:\\
\indent Firstly, we obtained the shape function using an embedding class- I approach under the redshift function $\phi(r)=-\frac{k}{r}$. We noticed that the obtained shape function follows the flare-out condition under asymptotically flatness conditions within the range $0<\delta<r_0$. Later, we inserted the shape function into the pressure components of the $4D$ EGB gravity. Further, we have investigated the energy conditions in the galactic halo with DM density profiles. We have observed that NEC is violated in the neighborhood of the throat. It was noticed the influence of GB coefficient $\alpha$ in the violation of NEC. Mathematically, one can check the RHS of Eqs. \eqref{4a9}, \eqref{4a12}, and \eqref{4a13} by simply putting the values of the parameters and taking the GB coefficient $\alpha>0$ provides a negative quantity. Also, we checked SEC for each DM profile and found that SEC was violated. Moreover, as we increase the value of $\alpha$, the contribution of violation becomes more. Additionally, DEC for both pressures was found satisfying near the throat. Moreover, one can check the summary of the energy conditions near the throat of the wormhole in Table-\ref{Table0}.\\
\indent Similar to the previous case, we extracted the shape function under the redshift function $\phi(r)=\frac{1}{2}\log(1+\frac{\eta^2}{r^2})$. We investigated the necessary criteria for a traversable wormhole, i.e., the flare-out condition, which is satisfied in the range $0<\lambda<r_0$. Later, we checked the NEC, SEC, and DEC for obtained solutions in the galactic halo regions. NEC is disrespected at the throat for $\alpha>0$. Moreover, one can check numerically from the expression given in Eq. \eqref{4b9}. DEC was investigated for each DM profile, and it was found that DEC was satisfied with both pressures in the galactic halos, except tangential DEC, which is violated in the URC DM profile. The calculated energy conditions are shown in Table-\ref{Table0}.\\ 
\indent Note that the confirming violated behavior of energy conditions supports the presence of the dark halos. We checked the behavior of energy conditions for each DM halo profile and noticed the violations of energy conditions, which means that traversable wormholes may exist in the galactic regions supported by dark matter in the context of $4D$ EGB gravity.\\
\indent Further, some physical features of wormholes, such as complexity factor, active gravitational mass, total gravitational mass, and embedding diagrams, have been explored in this paper. The complexity factor in the context of galactic DM halo wormholes in EGB gravity has been calculated, and it was noticed that the complexity factor converges to zero for high radial coordinates and GB coefficient $\alpha>0$. Such a study has been done in Ref. \cite{Butt3} for Casimir wormholes in higher dimensional EGB gravity. Moreover, the active gravitational mass for each DM density profile has been performed, and it has been observed that these DM models are physically acceptable under some restrictions. Rahaman et al. \cite{Rahaman7} studied the active gravitational mass of the NFW profile; however, in this paper, we have studied the active gravitational mass of URC and SFDM models along with the NFW model. Further, we have studied and numerically calculated the total gravitational energy for each DM profile under obtained shape functions. it was observed that $\mathcal{E}_g>0$, which signifies the presence of repulsion near the throat. This characteristic nature of $\mathcal{E}_g$ aligns with expectations for the formation of a physically viable wormhole.\\
In \cite{Atamurotov}, wormhole geometry in the galactic halos has been explored with two embedded wormhole-specific shape functions in the context of Einsteinian cubic gravity. They used observational data within the signature of the M87 galaxy and the Milky Way galaxy to check the effect of the dark matter halos. Further, in \cite{Rizwan3}, the author discussed wormhole solution with density profile obtained from modified Newtonian dynamics (MOND) with or without a scalar field in GR. Recently, wormhole solutions have been explored in $f(Q,T)$ gravity under different dark matter profiles such as URC and NFW profiles \cite{Sahoo}. In this paper, we have investigated wormhole solutions under URC, NFW, and SFDM profiles in $4D$ EGB gravity with two different redshift functions. Our study confirms that the obtained wormhole solutions might exist in the galactic halos within $4D$ EGB gravity. Also, traversable wormholes in the galactic halos with observational data sets in higher dimensional gravity would be an interesting problem that is being actively considered.

\section*{Data Availability}
There are no new data associated with this article.

\acknowledgments ZH acknowledges the Department of Science and Technology (DST), Government of India, New Delhi, for awarding a Senior Research Fellowship (File No. DST/INSPIRE Fellowship/2019/IF190911). PKS acknowledges the National Board for Higher Mathematics (NBHM) under the Department of Atomic Energy (DAE), Govt. of India, for financial support to carry out the Research project No.: 02011/3/2022 NBHM(R.P.)/R\&D II/2152 Dt.14.02.2022. We are very much grateful to the honorable referees and to the editor for the illuminating suggestions that have significantly improved our work in terms of research quality, and presentation.

\end{document}